\documentclass[twocolumn]{aastex631}

\usepackage{xcolor}
\usepackage{graphicx}
\usepackage{amsmath}
\usepackage{gensymb}
\usepackage{xcolor}

\begin{document}

\title{Dynamics and Energetics of Resistive, Thermally Conductive, and Radiative Plasma in Coronal Current Sheets due to Asymmetric External Perturbation}

\author[0009-0000-3578-8270]{Sripan Mondal}
\affiliation{Department of Physics, Indian Institute of Technology (BHU), Varanasi-221005, India. Email:- sripanmondal.rs.phy21@itbhu.ac.in}
\author{Akash Bairagi}
\affiliation{Department of Physics, Indian Institute of Technology (BHU), Varanasi-221005, India}
\author[0000-0002-1641-1539]{A.K. Srivastava}
\affiliation{Department of Physics, Indian Institute of Technology (BHU), Varanasi-221005, India. Email:- asrivastava.app@itbhu.ac.in}



\begin{abstract}
We study the asymmetric interaction of wave-like velocity perturbation with a coronal current sheet (CS) in the presence of resistivity, thermal conduction (TC) and radiative cooling (RC). We analyze the dynamics and energetics of CS in four cases, namely, (i) no energy loss, (ii) TC only, (iii) RC only and, (iv) TC+RC. Before fragmentation, thinning and elongation of the CS are found to be identical in all four cases and therefore independent of presence or absence of energy loss effects. Onset times, corresponding Lundquist numbers and aspect ratios suggest that TC advances the onset of fragmentation while RC has the opposite effect in comparison to absence of energy losses. Reconnection takes place at a higher rate in presence of TC and TC+RC in the tearing unstable CS. The speed of plasmoids are also found to be higher under the effect of TC and TC+RC.  In presence of TC and TC+RC, average density becomes higher within the tearing unstable CS than in other two cases. As expected, estimated average temperature  is increasing with highest and lowest rate in absence of energy losses and in presence of both TC and RC respectively.  After the onset of fragmentation, the rate of decrement of average magnetic energy density and increment of average kinetic energy density becomes higher in presence of TC and TC+RC  than in other two cases. Thus we conclude that presence of energy loss mechanisms critically influence the dynamics, energetics, and plasmoid formation within a reconnecting coronal CS.
\end{abstract}


\section{INTRODUCTION} 
Magnetic Reconnection forms the basis of the generation of kinetic energy of charged particles in plasma, and thermalize it by the liberation of the stored magnetic energy. Also, the magnetic reconnection facilitates as an essential mechanism to trigger the solar eruptions such as solar flares, coronal mass ejections (CMEs), etc. The occurrence of this process is not limited to the solar atmosphere only, whilst it eventually occurs in laboratory, space, magnetospheric and astrophysical plasma systems at diverse spatio-temporal scales \citep{2014masu.book.....P,2015A&ARv..23....4T,2022LRSP...19....1P}. Reconnection is basically a three dimensional process which can take place at the locations of high gradient of magnetic field such as magnetic nulls, separators, separatix surfaces, quasi-separators or hyperbolic flux tubes, bald patches etc \citep{2014masu.book.....P,2022LRSP...19....1P}. More importantly, irrespective of the initial magnetic geometry, formation of current sheet (CS) is essential for onset of the magnetic reconnection. However, CS can either remain stable or act as host of the magnetic reconnection. Either current dependent enhanced resistivity related to the magnetohydrodynamic turbulences \citep{2001EP&S...53..473S} or external perturbations in the form of EUV waves or shock waves \citep{2017JPlPh..83e2001V,2024ApJ...963..139M} can perturb the CS, which results in the magnetic fields present across the CS to reconnect or slip through each other to undergo several topological changes (See \citet{2014masu.book.....P,2022LRSP...19....1P} and references therein for more details). 

Even though, externally driven or forced reconnection has been studied theoretically for a long time \citep{1984SoPh...91..103S,1997AdSpR..19.1895O}, \citet{2019ApJ...887..137S} presented first direct imaging of the forced magnetic reconnection at considerably high rate at a temporary X-point in the solar corona as seen by Solar Dynamics Observatory (SDO)/Atmospheric Imaging Assembly (AIA) \citep{2012SoPh..275...17L}, where CS has been formed due to plasma inflows created by a prominence eruption. Subsequently in the follow-up observations by SDO/AIA, \citet{2021ApJ...920...18S} have studied extensively the forced magnetic reconnection which is creating the jet-like features and hot plasma flows in the off-limb corona. \citet{2020ApJ...905..150Z} have reported that a coronal loop system, connecting active regions (ARs) 11429 and 11430, consists of a null point undergoes a 40 minute oscillation when it is perturbed via shock wave which is further subjected to a magnetic reconnection occurred in the null point region.  Now, presence of plasma sheets, i.e., apparent CSs of diverse lengths and widths are ubiquitous during flares, eruptions, and jets \citep{2010ApJ...723L..28L,2018ApJ...853L..15L,2018ApJ...866...64C,2020A&A...644A.158P,2024MNRAS.528.1094Y,2024ApJ...974..104D}. Likewise, the generation and propagation of large scale EUV waves are also reported in several observational studies \citep{2013ApJ...776...58N,2014SoPh..289.4563M,2014SoPh..289.3233L,2015ApJ...812..173V,2017SoPh..292....7L,2018ApJ...868..107V,2018ApJ...861..105S,2023A&A...676A.144M}. Hence, in principle, both CSs and EUV waves can interact with each other in the solar corona. 

Although there are no direct coronal observations specifically focusing on interaction of EUV waves and the CS till now, recently \citet{2024NatCo..15.2667K} provided an observational evidence of the mode conversion of a fast MHD waves during its interaction with a 3D magnetic null. Symbiosis of waves and reconnection in the formed CS may play important role in augmenting existing knowledge about long-standing coronal heating problem \citep{2024arXiv241102180M,Sri24}.
However, their observational base-line should exist to further support the physical models and their exclusive physical scenario. Moreover, there are observations of flare or eruption initiated via shock waves or EUV waves originated during another distant earlier flare or eruption \citep[e.g.,][and so on]{1981sfmh.book...47S,1983SoPh...89..355F,2001ApJ...559.1171W}. Therefore, such spatio-temporally correlated sympathetic events may serve as spatially-unresolved potential indirect signature of magnetic reconnection initiated via interaction of EUV or shock waves with magnetic null or CS. Moreover, in other plasma systems such as in Earth's magnetosphere, it has been reported that non-reconnecting CS got compressed due to its interaction with bow shock near to magnetopause in the magnetosheath which can result in onset of magnetic reconnection within that CS \citep{2007JGRA..11212219M,2021ApJ...913..142K}. Therefore, more focused observations on this particular aspect of interaction of EUV waves or fast shock waves and non-reconnecting quasi-stable CSs may provide some direct signature of such interaction also in the solar corona in future. 

Most of the large scale impulsive eruptions such as flares, CMEs, filaments are attributed to the fast reconnection. Fragmentation of elongated and narrow CS and subsequent plasmoid formation via tearing mode instability has been extensively studied for decades as one of the mechanisms to achieve high reconnection rate theoretically  \citep{2007PhPl...14j0703L,2009PhPl...16k2102B,2010PhPl...17f2104H,2016ApJ...818...20H,2017ApJ...850..142C}. High resolution observational facilities further confirms the presence of these fragmented structures and dense plasma blobs within elongated narrow intensity features, i.e., CSs or even in jets \citep{2007ApJ...655..591R,2013ApJ...771L..14G, 2016SoPh..291..859Z,2016ApJ...826...94K,2017ApJ...841...49C,2018ApJ...866...64C,2020A&A...644A.158P,2020ApJ...900..192F,2021ApJ...922..117F}. Although, tearing instability has been studied extensively in solar corona, there are only a few studies where this tearing mode has been achieved without any localized enhancement of resistivity \citep{2009PhPl...16a2102B,2022arXiv220900149S,2024A&A...683A..95F}. Also, the role of various energy loss mechanisms such as thermal conduction (TC) and radiative cooling (RC) in CS dynamics and associated magnetic reconnection is not explored in greater details. Even though both TC and RC are energy loss processes, the basic mechanisms are little different such as- (i) TC depends on the temperature gradient, i.e., it conducts heat from high temperature regions to relatively cooler regions, whereas (ii) RC is directly connected to the density of the emitter and also on an complex temperature dependent cooling functions. Therefore, it is important to explore their individual as well as resultant effects on the dynamical and more importantly thermal characteristics related to the magnetic reconnection. The observable apparent current sheets are unique in solar corona, which manifests the evolution of temperature, density, and plasma flows in them distinctly \citep{2024ApJ...974..104D}. This is most possibly attributed due to the different roles of the thermodynamical properties set within them.

\citet{2021SoPh..296...74L,2021SoPh..296...93L} reported that presence of non-adiabatic effects such as radiative loss, thermal conduction etc can modify growth rate of tearing instability in the linear regime. \citet{2022A&A...666A..28S} explored the role of radiative loss and background heating on tearing instability in a CS via a 2D resistive MHD simulation. They found that explosive nature of reconnection is more achievable at relatively lower Lundquist number due to the positive feedback between tearing and thermal instabilities caused by the presence of radiative loss. But, they used magnetic field perturbations to achieve the magnetic reconnection in absence of TC which is an effective energy loss effect in the solar corona. Besides, they did not consider presence of guide field which is omnipresent in solar corona especially in active regions and can suppress plasmoid formation. \citet{2024ApJ...963..139M} studied the external forcing aspect in the form of velocity perturbation for the onset of reconnection and subsequent plasmoid formation in absence of any kind of energy loss effects. Absence of any energy loss mechanism results in temperature enhancement upto 20 MK at the core of plasmoids. So, it is interesting to see the effects of energy loss mechanisms on the temperature of plasmoids and the associated CSs, and how these candidates overall influence the physical processes and morphological appearance. Also, it is of utmost importance to check the temporal evolution of different forms of energy such as magnetic, kinetic energy etc for different cases, i.e., without any energy loss effects, with TC only, with RC only and in presence of both TC and RC.

In this paper, we follow \citet{2024ApJ...963..139M} to have an external forcing, however, in a new form of anisotropically propagating velocity pulse for perturbing the model current sheet in the presence of thermal conduction and radiative cooling. We explore the evolution of different forms of energies with time in aforementioned different cases and make possible physicswise interpretations. In Sect 2, we describe the numerical setup and methods. In Sect 3, we report the results. In Sect 4, we summarize and compare the results with previous studies along with their possible physical implications.

\section{Numerical Setup and Methods} 
We perform a 2.5 dimensional magnetohydrodynamic simulation to study the onset, evolution of the magnetic reconnection, and associated energetics in a force-free Harris CS as given by
\citep{2018ApJS..234...30X,2024ApJ...963..139M}:
\begin{equation}
   B_{x} = 0
\end{equation}
\begin{equation}
   B_{y} = - B_{0}~\tanh \left(\frac{x}{l}\right)
\end{equation}
\begin{equation}
   B_{z} = B_{0}~{\rm sech} \left(\frac{x}{l}\right), 
\end{equation}
where \(B_{x}\) and \(B_{y}\) are the magnetic field components in the plane in which entire dynamical and energetic processes of CS are captured. Non zero \(B_{z}\) component ensures vanishing Lorentz force to maintain the unstratified homogeneous corona in magnetohydrostatic equilibrium initially in presence of uniform plasma pressure. Basically, we use the same initial magnetic field configuration to the one as described in \citet{2024ApJ...963..139M}, i.e., the magnetic field amplitude ($B_{0}$) and CS half-width ($l$) are 10 G and 1.5 Mm, respectively. Other physical conditions  also resemble to the typical corona similar to the reported values as given in \citet{2024ApJ...963..139M}. For example, the uniform temperature and density are respectively 1 MK and \(2.34 \times 10^{-15}\) g \(\mathrm{cm^{-3}}\) with a plasma \(\beta\) of 0.079. Basically, in \citet{2024ApJ...963..139M}, they studied the onset of magnetic reconnection and tearing mode instability and subsequent multiple stages of plasmoid formations in details when the CS has been perturbed symmetrically at the centre of its length by an isotropically propagating velocity pulse. But in reality, it is more likely that any perturbation will be anisotropic and it can interact with the CS at any point along its length. Also, it is interesting to consider energy loss mechanisms such as field-aligned TC and RC to study their individual and collective effects on the entire physical processes and dynamics of CS. 

Since, we are studying the dynamics in a 2.5D space, therefore, physical variables should not possess any z-dependence, i.e., any perturbation also must be function of x and y only. So, we perturb the initial equilibrium with a force-free stable CS via imposing an anisotropically propagating Gaussian pulse given by \citep{2024arXiv241102180M}
\begin{equation}
    v_{x} = v_{0}~\mathrm{exp} \left(- (\frac{x-x_{0}}{w_{x}})^{2}-(\frac{y-y_{0}}{w_{y}})^{2}\right).
\end{equation}
Here $x_{0}$ is set at 15 Mm left from the centre of the CS at $x=0$, and $y_{0}$ is set at a height of 65 Mm. Therefore, the perturbation will interact with the CS at y= 65 Mm with the highest amplitude due to lowest distance between its epicentre and the CS. This further confirms an interaction that is asymmetric about the length of the CS. The widths of this Gaussian pulse in X and Y direction ,i.e., $w_{x}$ and $w_{y}$ are taken to be 6 Mm and 2 Mm respectively. The amplitude ($v_{0}$) of the velocity pulse is 350 km \(\mathrm{s^{-1}}\) that is $0.6\mathrm{v_{A}}$, where $v_{A}=580\ \mathrm{km\ s^{-1}}$ is the Alfv\'en speed for our choice of the parameters. It is to be noted that the amplitude $v_{0}$ is kept same as in \citet{2024ApJ...963..139M}. This pulse is basically mimicking a large scale EUV wave often observed to be propagating in the large scale corona after its transient generation during onset of solar flares and coronal mass ejections (CMEs) \citep[e.g.,][]{2014MNRAS.444.1119Z,2018ApJ...868L..33L,2022ApJ...929L...4Z}. 

We use open source MPI-AMRVAC \footnote{\url{http://amrvac.org}} \citep{2018ApJS..234...30X,2023A&A...673A..66K} to numerically solve the following resistive magnetohydrodynamic (MHD) equations in their conservative form \citep{2022A&A...666A..28S,2023A&A...678A.132S,2024arXiv241102180M}:
\begin{equation} 
\frac{\partial \rho}{\partial t} + \vec{\nabla} \cdot ( \rho \vec{V} ) = 0,
\end{equation}

\begin{equation}
  \frac{\partial}{\partial t}(\rho \vec{v}) + \vec{\nabla} \cdot \left [ \rho \vec{v}\vec{v}  + p_{tot}\vec{I} - \frac{\vec{B}\vec{B}}{4\pi} \right ] = 0 ,
\end{equation}

\begin{equation}
\begin{split}
\frac{\partial e}{\partial t} +  \vec{\nabla} \cdot \left( e\vec{v} + p_{tot}\vec{v} -\frac{\vec{B}\vec{B}}{4\pi} \cdot \vec{v}\right)  = \eta \vec{J^{2}}-
   \vec{B} \cdot \vec{\nabla} \times (\eta \vec{J})\\
   +\vec{\nabla}_{\parallel} \cdot (\kappa_{\parallel} \vec{\nabla}_{\parallel} T)-\rho^{2}\Lambda(T),
\end{split}   
\end{equation}

\begin{equation}
  \frac{\partial \vec{B}}{\partial t} + \vec{\nabla} \cdot \left(\vec{v}\vec{B} - \vec{B}\vec{v}\right)+ \vec{\nabla} \times (\eta \vec{J}) = 0,
\end{equation}

\quad \textrm{where} \quad
\begin{equation}
\begin{split}
p_{tot}= p+ \frac{\vec{B}^2}{8\pi}, ~~e = \frac{p}{\gamma-1} + \frac{1}{2}\rho v^{2} + \frac{B^2}{8\pi}, 
\\~~\vec{J} = \frac{\vec{\nabla} \times \vec{B}}{4\pi},
~~\vec{\nabla} \cdot \vec{B} =0.
\end{split}
\end{equation}
Here, magnetic diffusivity $\eta$ is taken to be \(2.4 \times 10^{8}~\mathrm{m^{2}s^{-1}}\) throughout the simulation domain, which corresponds to a Lundquist number of \(4.8 \times 10^{5}\). It is to be noted that this diffusivity is kept same as in \citet{2024ApJ...963..139M}. $\kappa_{\parallel}=10^{-6}~T^{5/2}~\mathrm{erg~cm^{-1}~s^{-1}~K^{-1}}$ is the component of the thermal conduction tensor along the magnetic field. It is worth to mention that since solar corona is fully ionised and magnetically dominated, we do not consider the thermal conduction perpendicular to the magnetic field. But since, we are interested in formation and evolution of the plasmoids in the present work, presence of perpendicular thermal conduction may govern the formation and thermal characteristics of such fine structures \citep{1991SoPh..134..247V}. However, \citet{1992SoPh..142..265I} suggested that presence of finite resistivity results in similar effects as presence of perpendicular thermal conduction in formation and evolution of fine structures such as plasmoids etc. Therefore, absence of perpendicular thermal conduction will not change the characteristics of the reconnection dynamics and plasmoids in presence of the finite resistivity. The optically thin radiative cooling depends on the local density, and the temperature-sensitive cooling models. In this work, we use the cooling model as reported by \citet{2008ApJ...689..585C}.

\begin{figure*}
\centerline{\hspace*{0.013\textwidth}
         \includegraphics[height=5 cm,trim={0 0 0 0},clip]{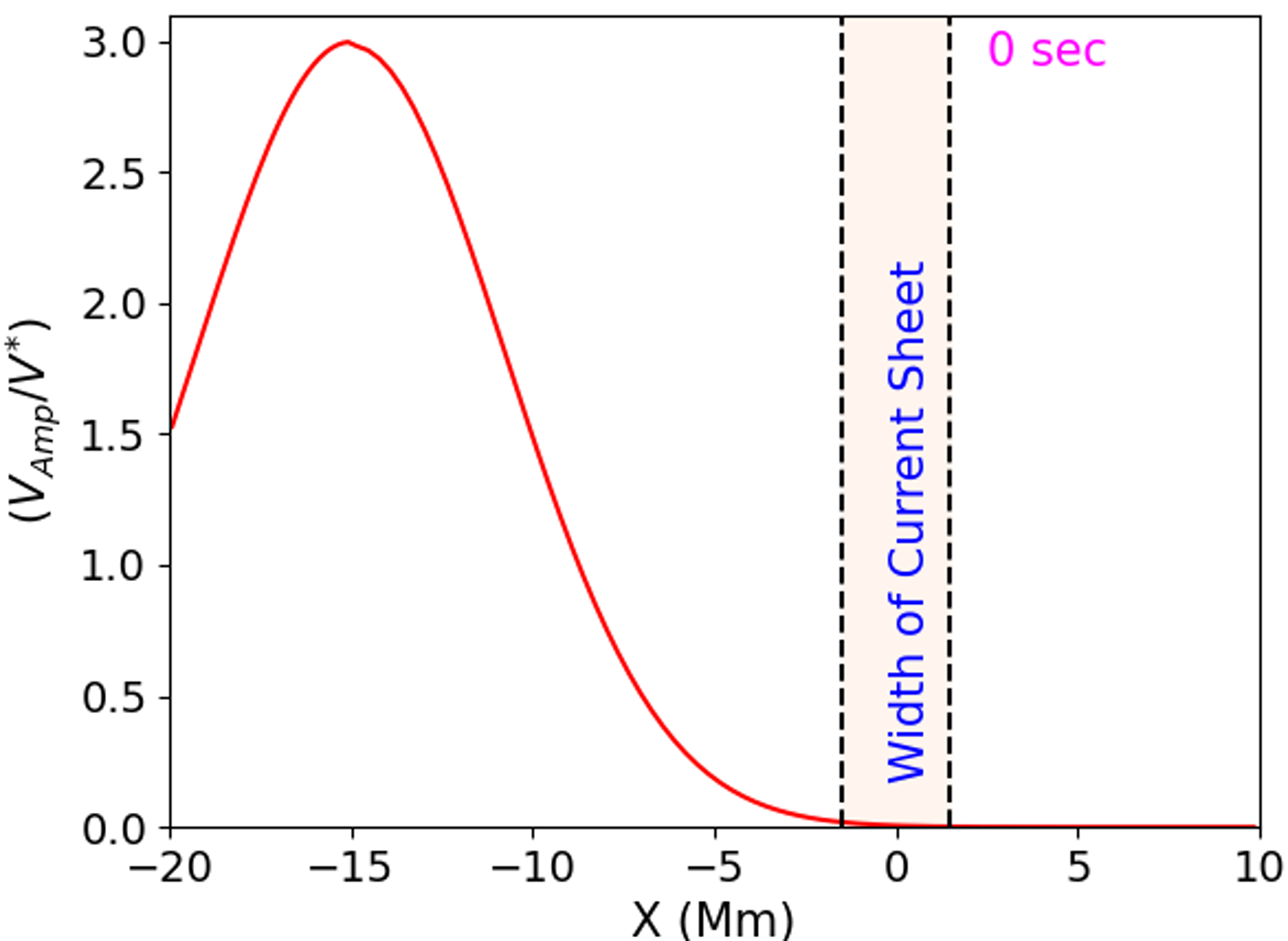}
         \hspace*{-0.01\textwidth}
         \includegraphics[height=5 cm,trim={0 0 0 0},clip]{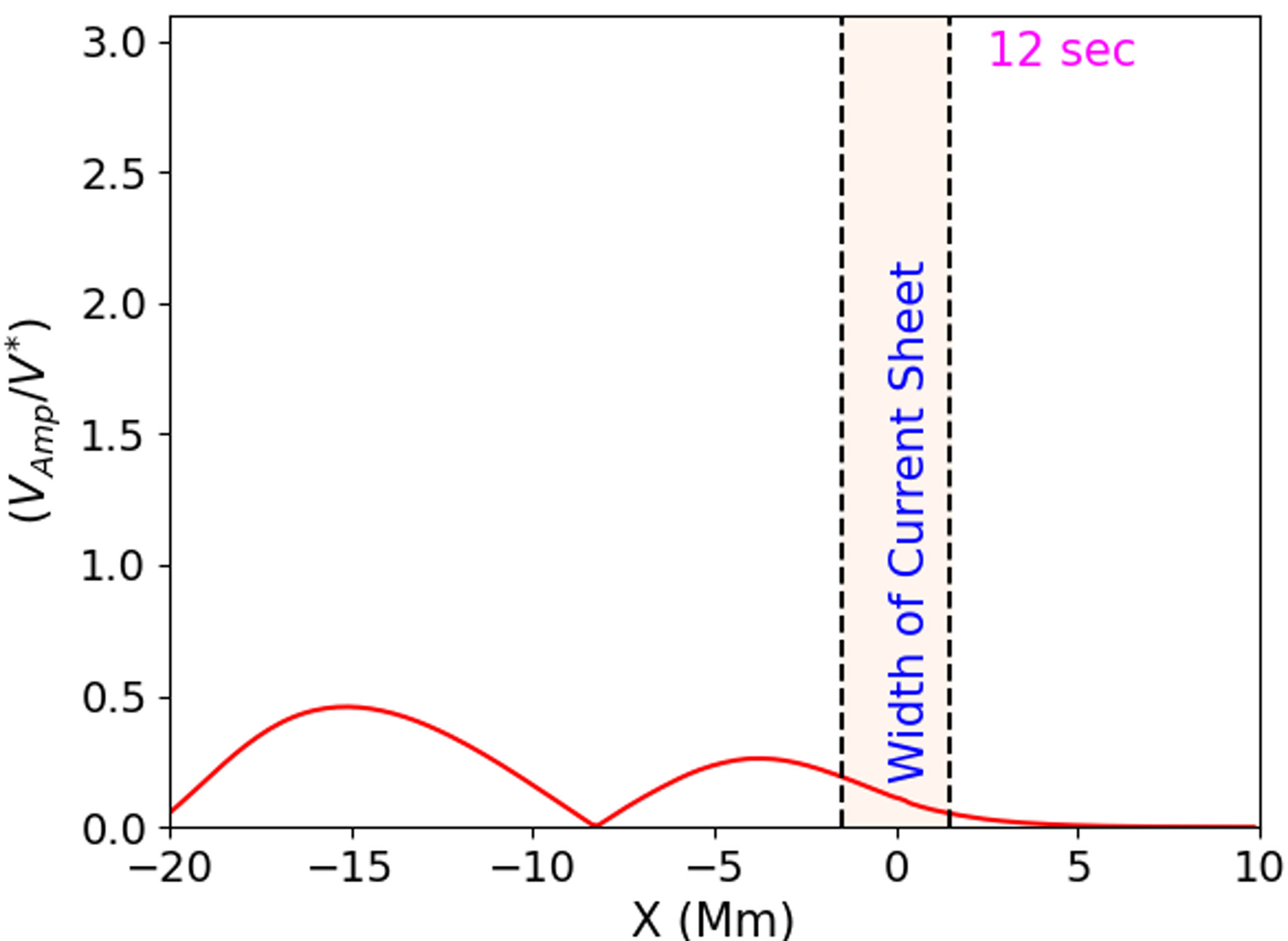}
         \hspace*{-0.01\textwidth}
         \includegraphics[height=5 cm,trim={0 0 0 0},clip]{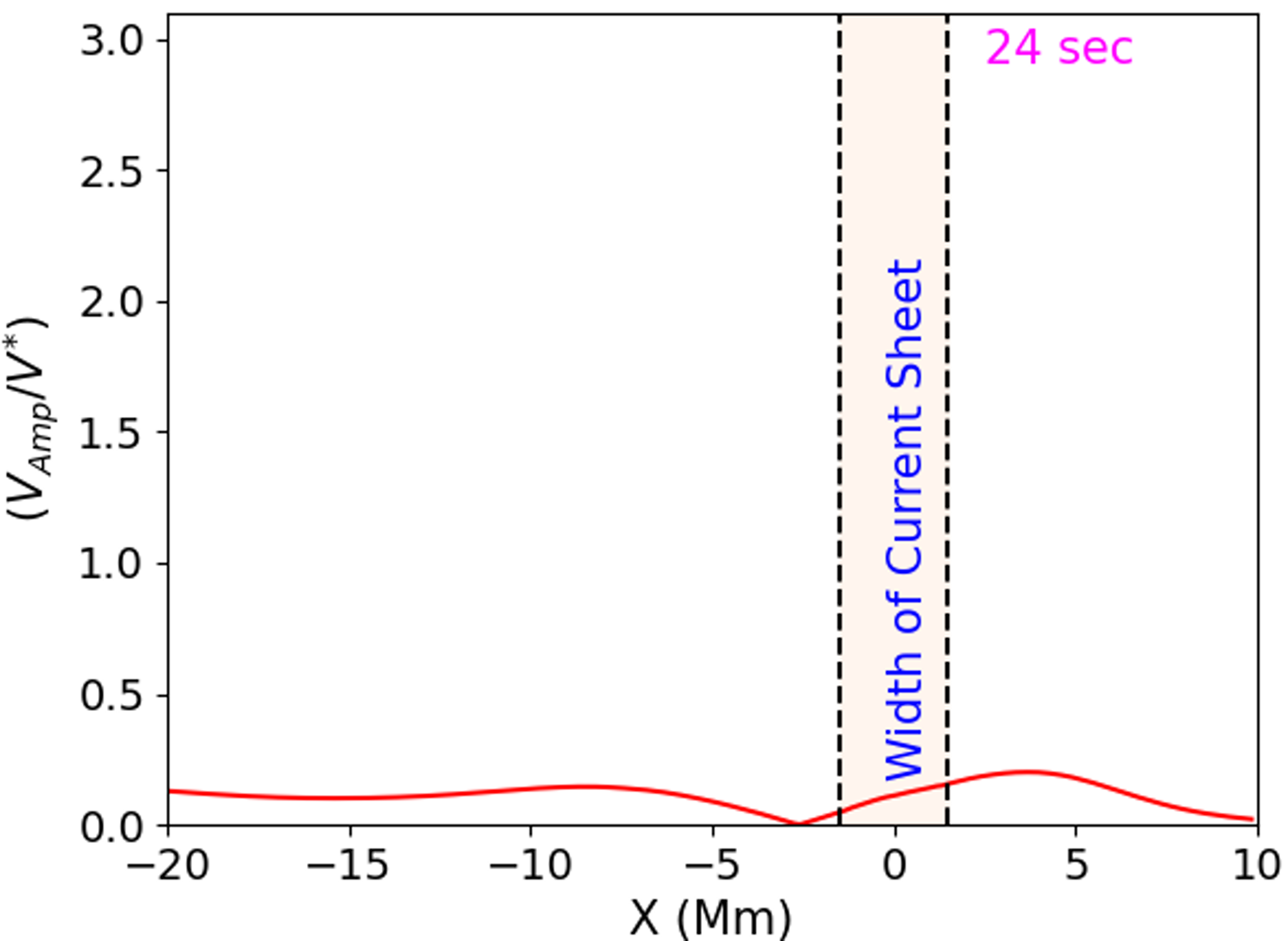}
         }
\vspace{-0.40\textwidth}
\centerline{ \large \bf      
   \hfill}
\vspace{0.37\textwidth}    
\caption{Temporal evolution of the initial velocity disturbance during its passage towards and through the CS as estimated at y = 65 Mm. At 0 s, the disturbance is properly Gaussian shaped centred at x = - 15 Mm. As it propagates, it gets distorted and its amplitude becomes much smaller at the instance of interaction with the current sheet (12 s). The shaded region within vertical dashed black lines show the initial location and width of the CS in all the panels. At 24 sec, the initial perturbation almost passed through the CS. It is to be noted that the velocity disturbance evolves exactly in similar manner with same amplitude before or even after its interaction with the CS in all four cases such as no energy loss (NEL), TC only, RC only and TC+RC. Hence, we do not explicitly show those overlapping profiles in this figure.}
\label{label 1}
\end{figure*} 
Since the above mentioned Equations (1)-(9) are solved numerically, all of the physical variables are made dimensionless using the corresponding factors as follows \citep{2017ApJ...841..106Z,2024ApJ...963..139M}: \(L^{*}= 10^{9}\) cm, \(\rho^{*}= 2.34 \times 10^{-15}\) g \(\mathrm{cm^{-3}}\), \(V^{*} = 116.45\) km \(\mathrm{s^{-1}}\), \(P^{*} = 0.32\) dyne \(\mathrm{cm^{-2}}\), \(B^{*}\)= 2 Gauss, \(T^{*}\)= 1 MK and \(J^{*}\)= 4.77 statA~cm$^{-2}$. The simulation domain spatially extends from -100 Mm to 100 Mm in the $x$-direction and 0 to 200 Mm in the $y$-direction. The physical total time duration of our simulation is 962 seconds (i.e., around 16 minutes). The initial coarse spatial resolution is 1.25 Mm in both the directions which reaches to highest resolution of 78 km after  adaptive mesh refinement (AMR) of four-levels. Temporal integration is carried out using `twostep' Runge-Kutta method. Estimation of the flux at cell interfaces is performed via `Harten Lax vanLeer (HLL)' flux scheme \citep{1983JCoPh..49..357H}. All the physical variables are zero-order extrapolated using the closest inner mesh cell value to all of the ghost cells, i.e., the gradient is kept zero for all the variables across each boundary to ensure no reflection.

\begin{figure*}
\centerline{\hspace*{0.02\textwidth}
         \includegraphics[width=0.28\textwidth,clip=]{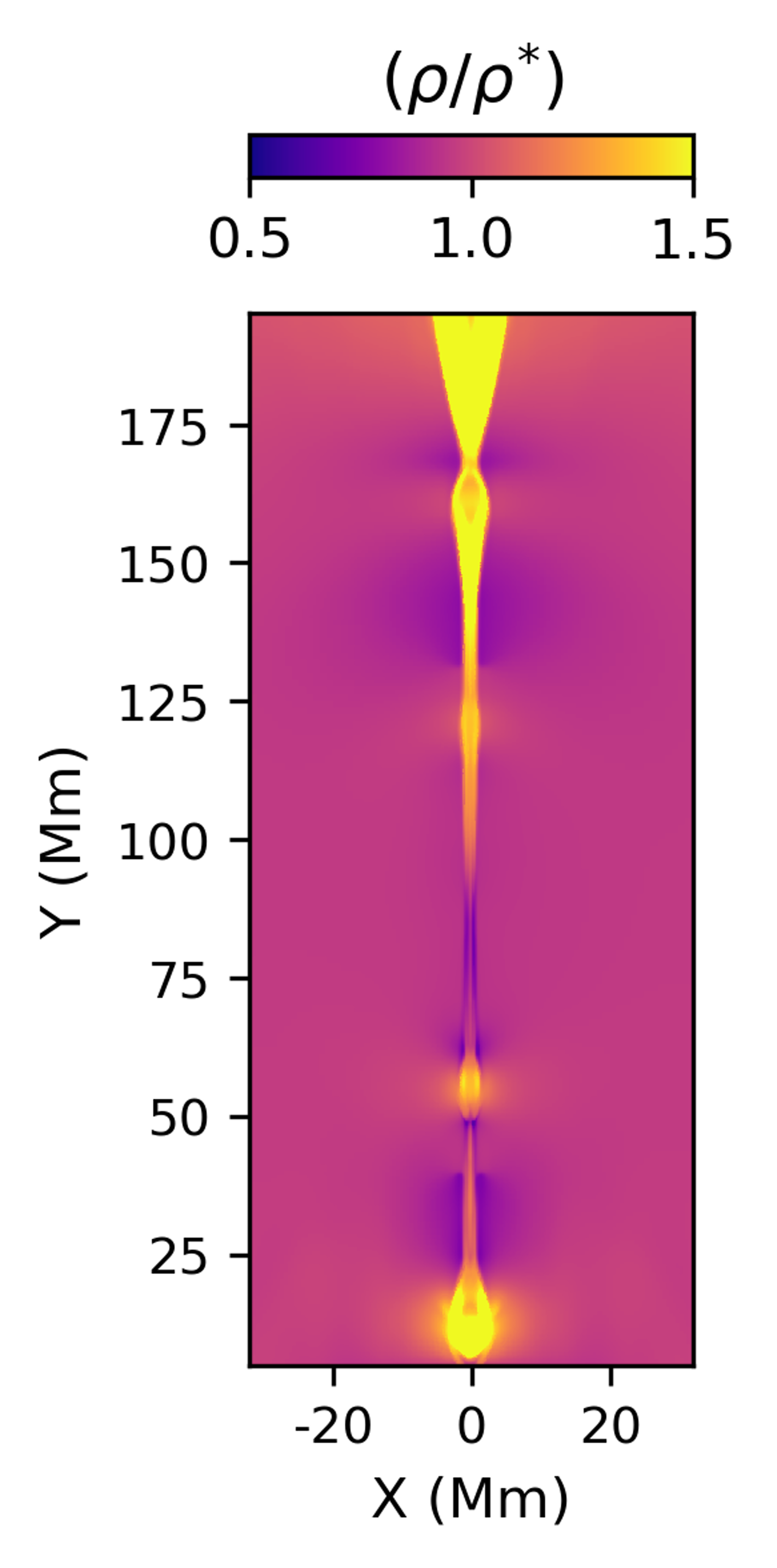}
         \hspace*{-0.03\textwidth}
         \includegraphics[width=0.28\textwidth,clip=]{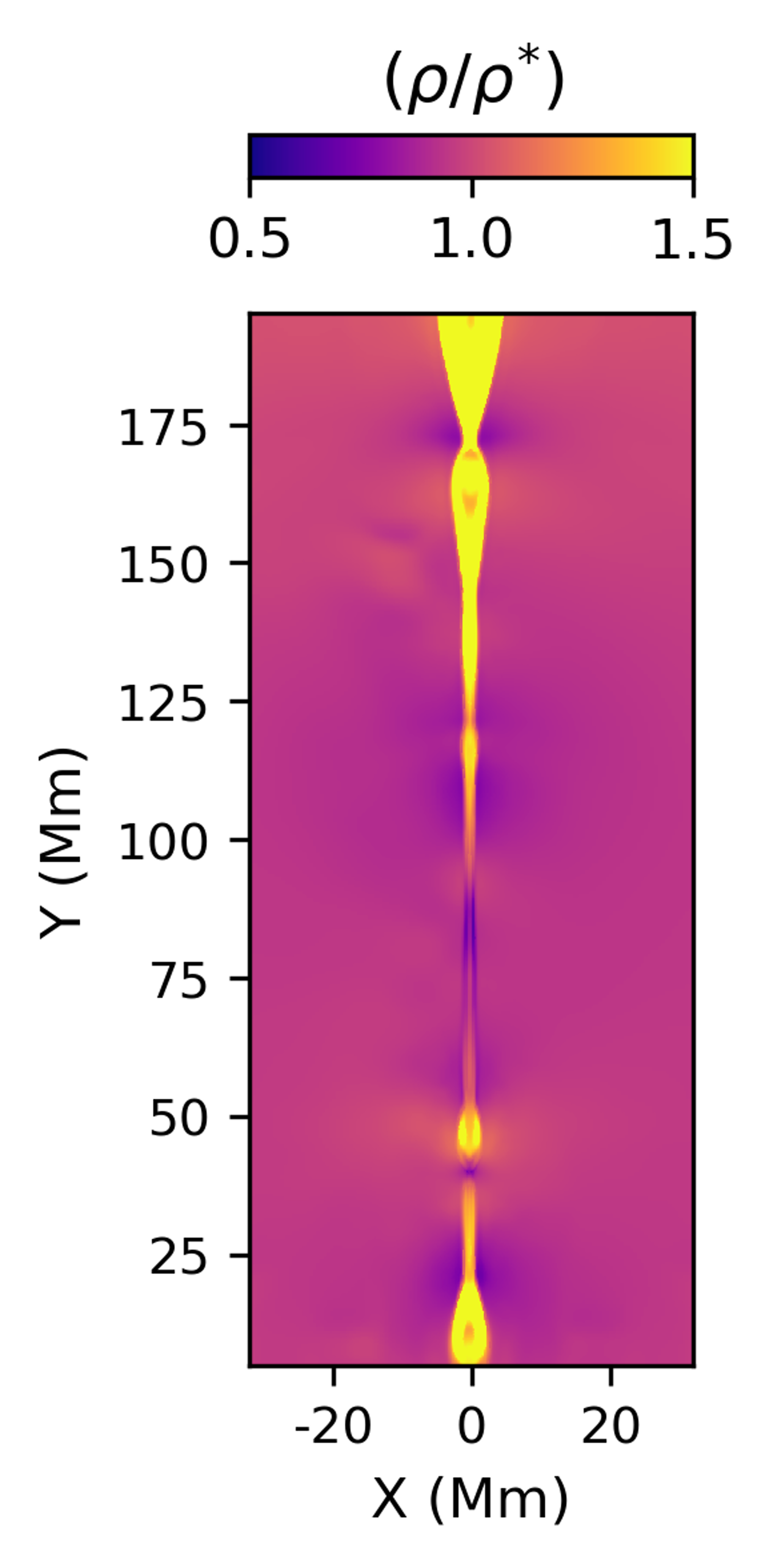}
         \hspace*{-0.03\textwidth}
         \includegraphics[width=0.28\textwidth,clip=]{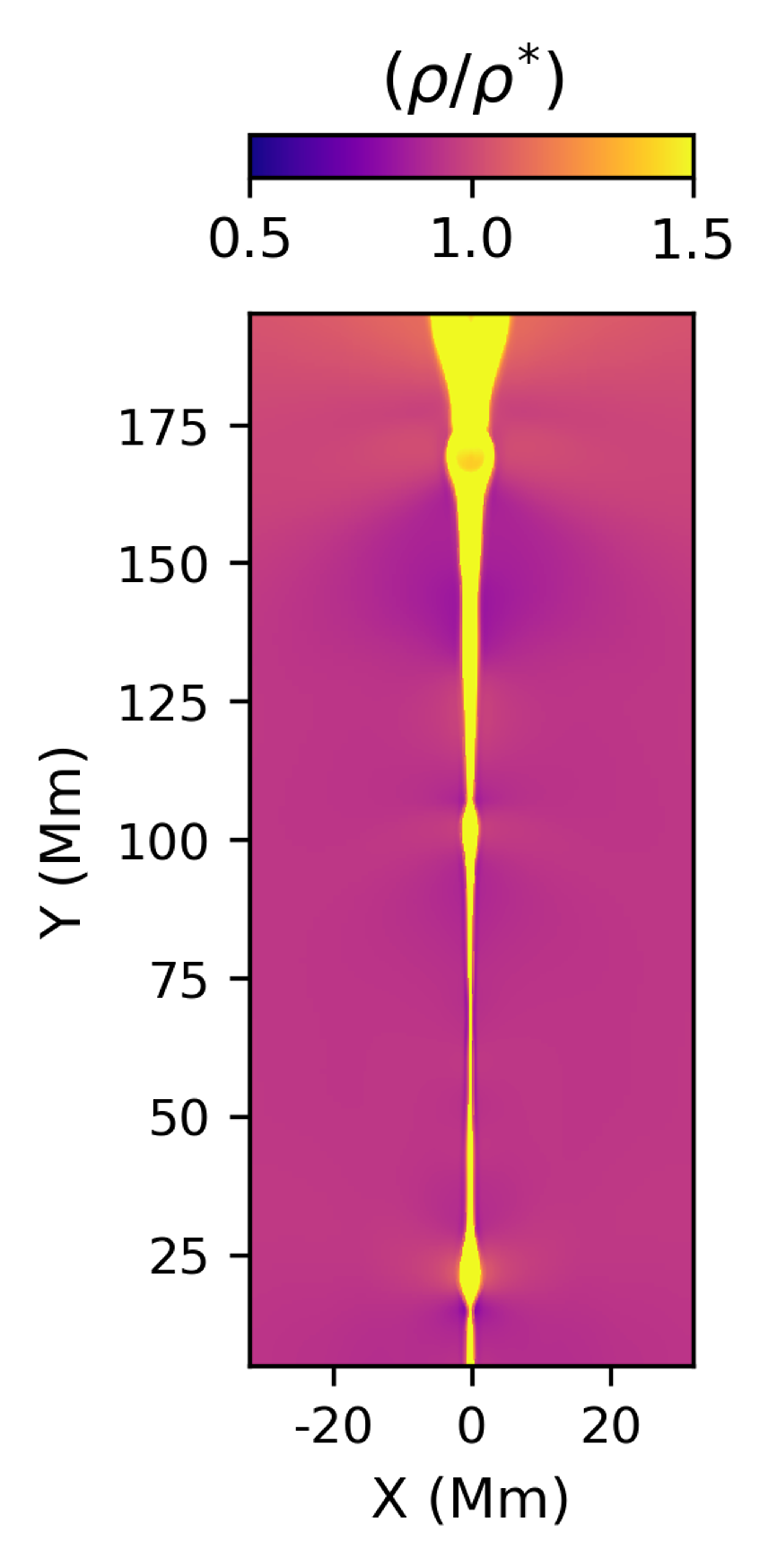}
         \hspace*{-0.03\textwidth}
         \includegraphics[width=0.28\textwidth,clip=]{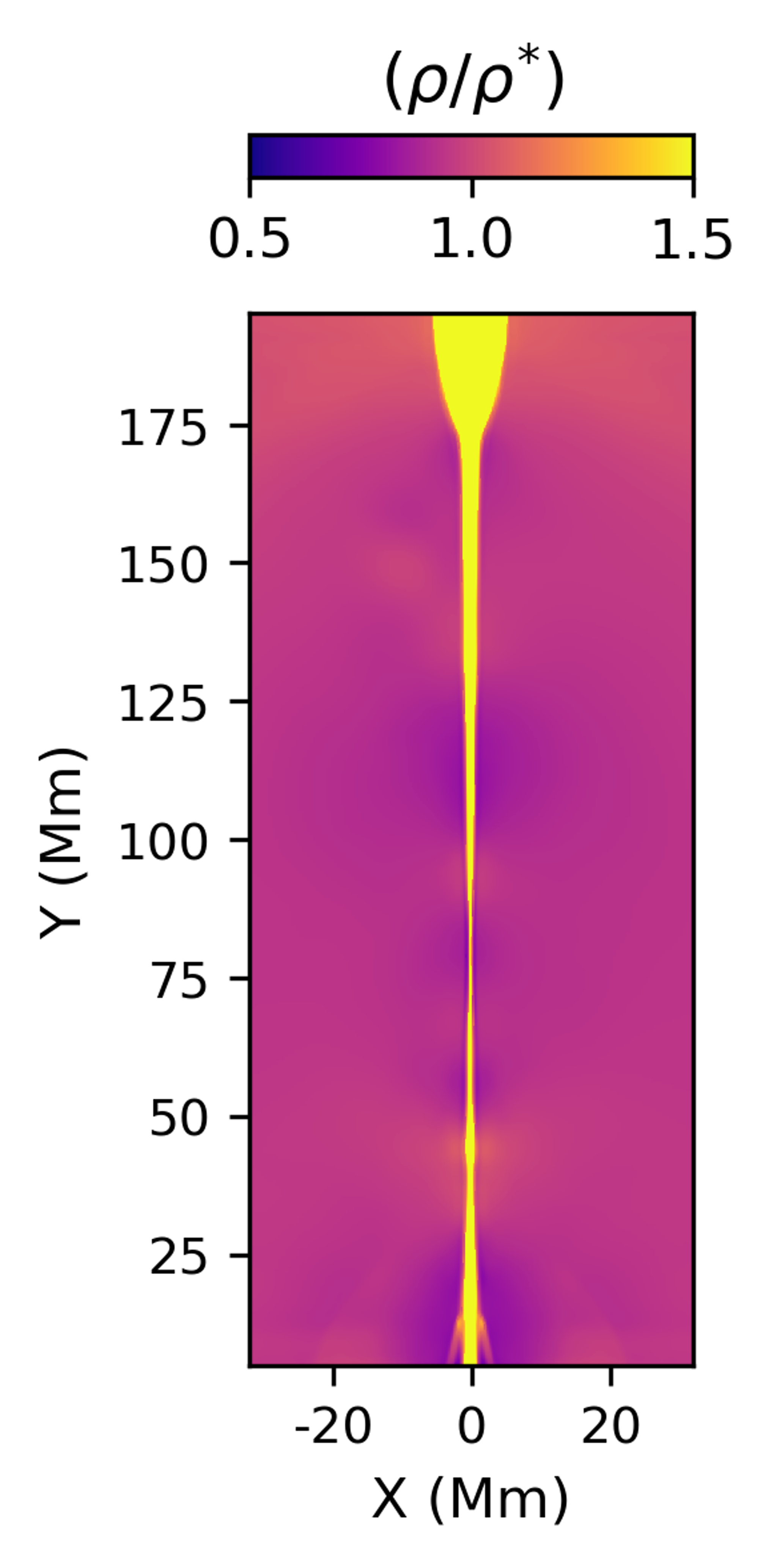}
        }
\vspace{-0.59\textwidth}   
\centerline{ \Large     
                          
\hspace{0.10 \textwidth}  \color{black}{NEL}
\hspace{0.16\textwidth}  \color{black}{RC only}
\hspace{0.16\textwidth}  \color{black}{TC only}
\hspace{0.15\textwidth}  \color{black}{TC+RC}

   \hfill}
\vspace{0.55\textwidth}    
\centerline{\hspace*{0.008\textwidth}
         \includegraphics[width=0.28\textwidth,clip=]{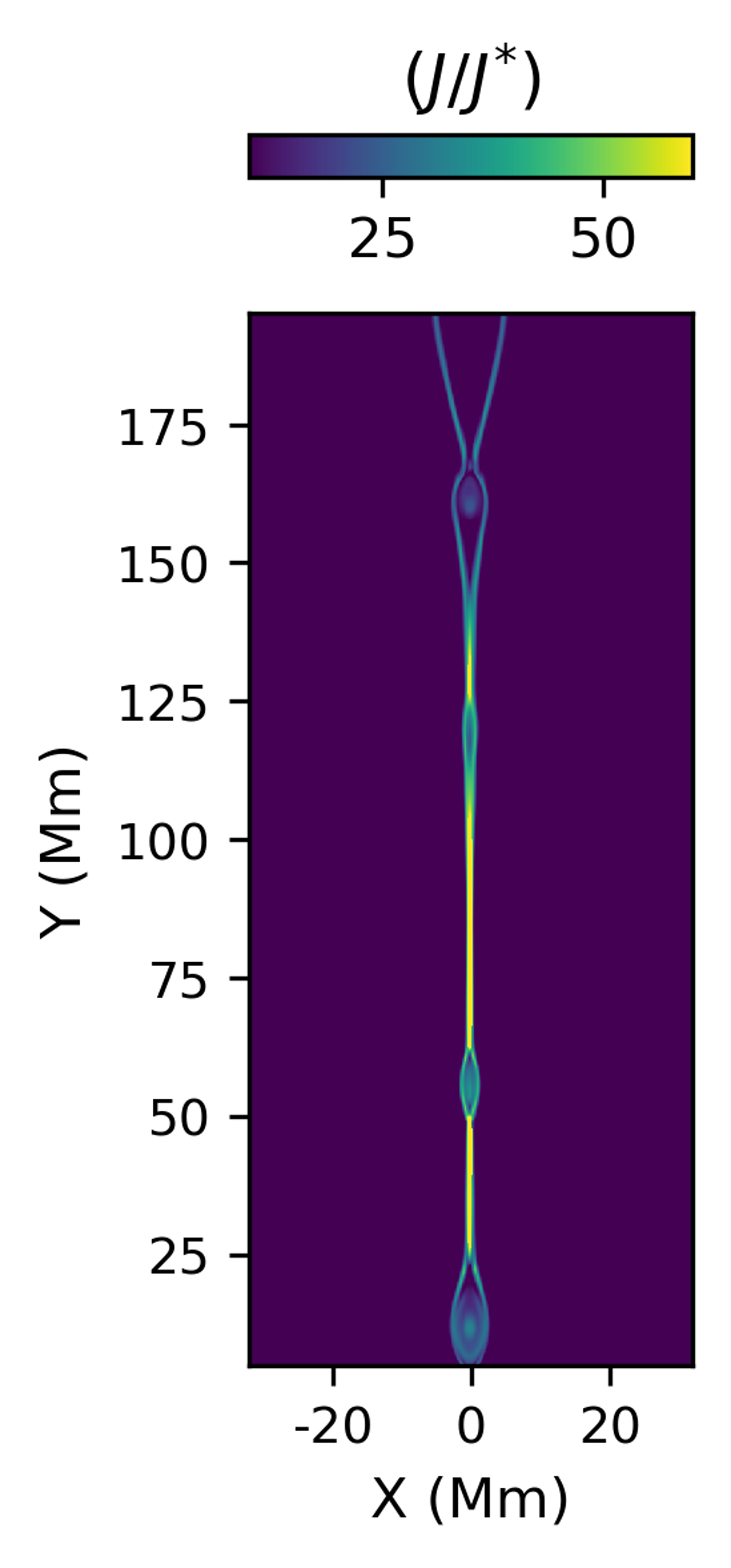}
         \hspace*{-0.03\textwidth}
         \includegraphics[width=0.28\textwidth,clip=]{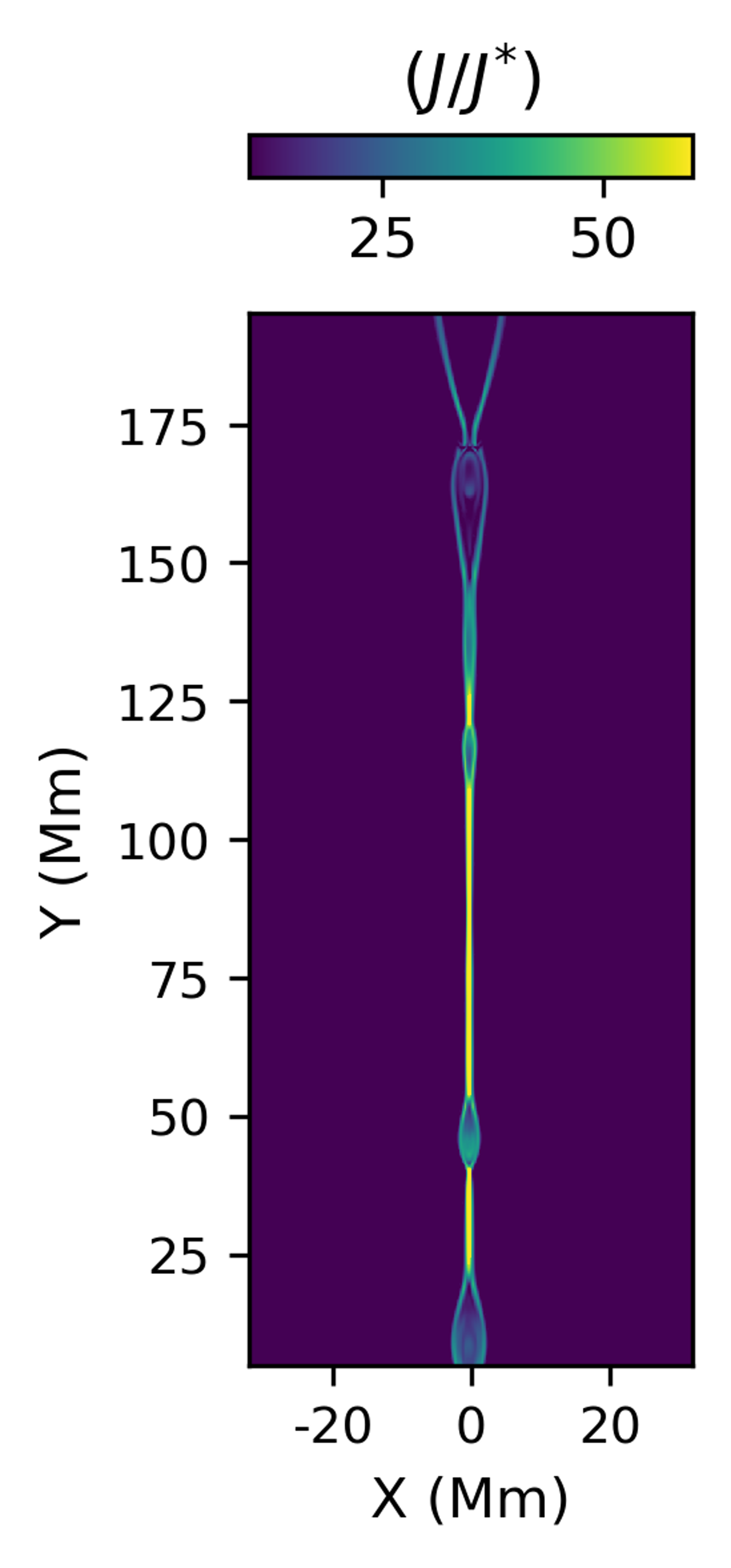}
         \hspace*{-0.03\textwidth}
         \includegraphics[width=0.28\textwidth,clip=]{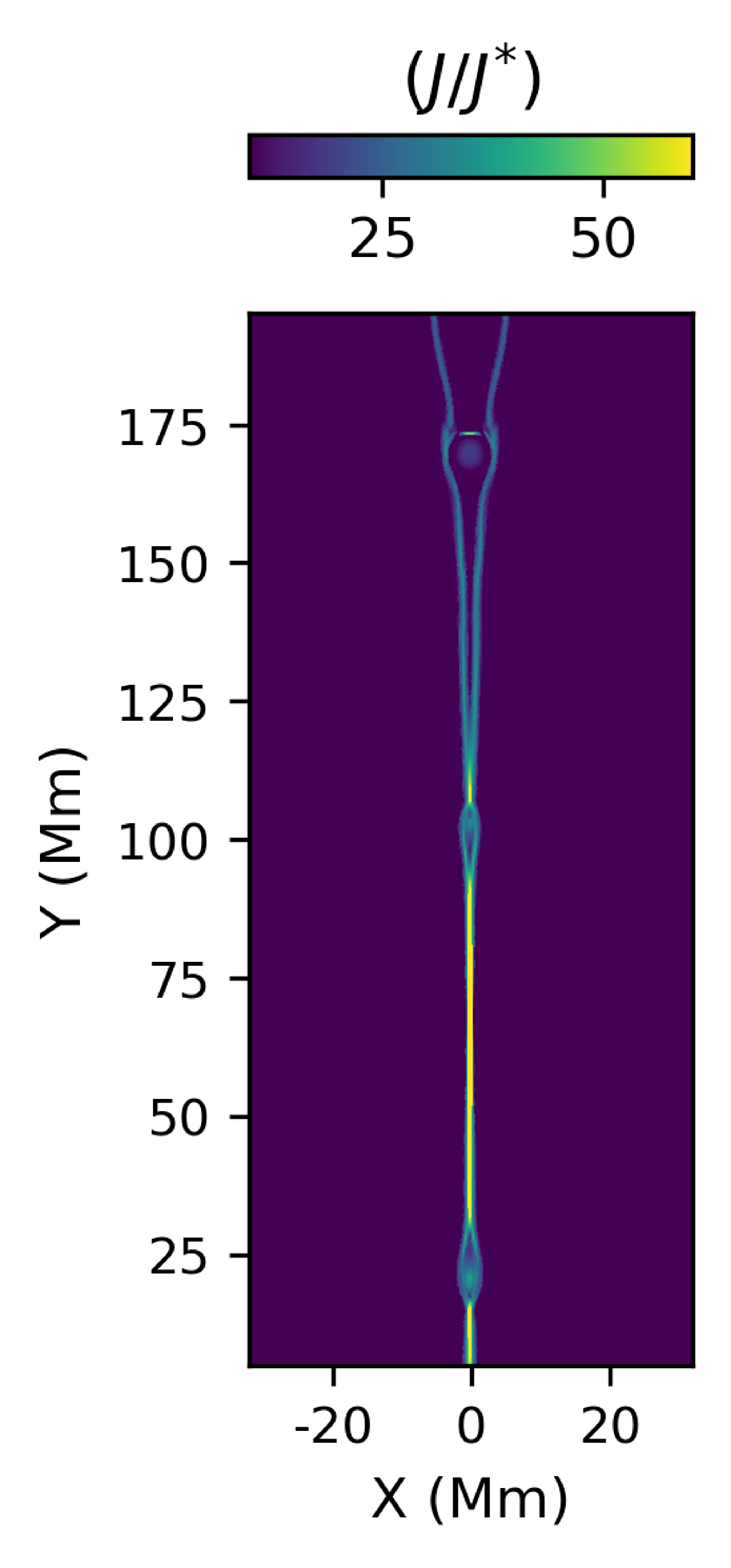}
         \hspace*{-0.03\textwidth}
         \includegraphics[width=0.28\textwidth,clip=]{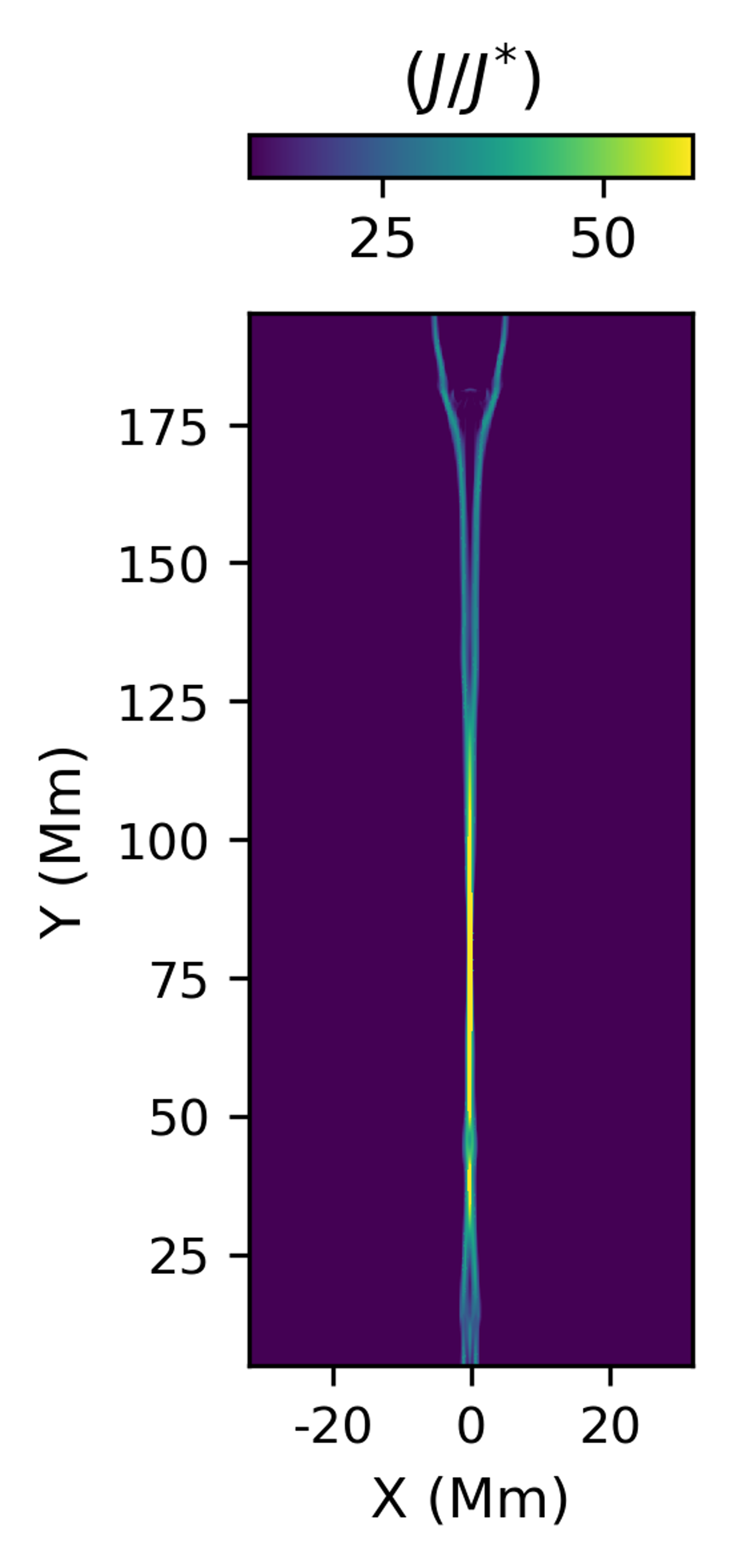}
        }
\vspace{0.04\textwidth} 
\caption{Top: Spatial density distribution within the tearing mode unstable current sheet and its immediate surroundings at 950 s in absence of energy loss effects (denoted as NEL), presence of RC, presence of TC, and presence of TC+RC (left to right). Bottom: Spatial distribution of current density there at 950 s for the aforementioned four cases. An animated version of the entire dynamics from 240 s to 962 s is available in the online version. The real time duration of the animation is 6 s. The triggering and initial evolution of the CS is similar to \citet{2024ApJ...963..139M} except the point of initial interaction being at $y = 65~\mathrm{Mm}$ instead of $y = 100~\mathrm{Mm}$, i.e., asymmetric about its length. Hence, we do not present snapshots from 0 s to 228 s explicitly.}
\label{label 2}
\end{figure*}
\section{RESULTS}
We follow \citet{2024ApJ...963..139M} to initiate thinning of the CS due to inward pressure and density gradient, which result in via an interaction of velocity perturbation with the CS and its immediate surroundings (See Figure 1 and section 3.1.1 of \citet{2024ApJ...963..139M} for more details). However, on contrary to \citet{2024ApJ...963..139M}, we perturb the CS asymmetrically about its length. Even though the initial amplitude of velocity perturbation is taken to be $3V^{*}$, i.e., 350 km \(\mathrm{s^{-1}}\) (See left panel of Figure~\ref{label 1}) , we find that the resultant amplitude of this perturbation is around $0.25V^{*}$, i.e., 30 km \(\mathrm{s^{-1}}\) during its interaction with the CS (See middle panel of Figure~\ref{label 1}). We do not find any sawtooth like profile as an indication of shock formation from this initial disturbances as in \citet{2024arXiv241102180M}. Rather they propagate as sinusoidal features such as waves in the present case. Since, the leading peak travels a distance of roughly 11 Mm in first 12 s, the estimated speed of the disturbance is around 915 km \(\mathrm{s^{-1}}\). Hence, the velocity disturbance works as a fast magnetosonic wave propagating across vertical field lines. In the present work, our main aim is to find out the differences in the entire dynamics due to presence of energy loss effects in the form of field-aligned TC and RC. So, we will discuss the differences and/or similarities in the stage-wise dynamics and thermal characteristics for four different cases such as--[I] No energy loss (hereafter, NEL); [II] RC only; [III] TC only; and, [IV] TC+RC. It is worth to mention that presence or absence of energy loss effects such as TC and RC do not affect the evolution of the velocity perturbation before or even after its interaction with the CS. In addition, the perturbation in each plasma parameters such as density, temperature etc caused due to the velocity disturbance in all four cases do not differ from each other significantly. Therefore, we do not exhibit any explicit presentation of overlapping profiles of propagating time-evolved velocity disturbance or perturbation in plasma parameters in above mentioned four cases in Figure~\ref{label 1}. Since the velocity perturbation takes roughly 24 s only to traverse 15 Mm while reaching the CS in the homogeneous ambient corona, it does not get enough time to dissipate within the region of interest due to presence of energy loss effects as this time scale is very less compared to the typical thermal conductive and radiative cooling time scales. However, the visually detectable differences in density and current density snapshots at 950 s for four different cases indicate that energy loss effects may affect the dynamics in the time-evolving CS (See Figure~\ref{label 2} and associated animation). So, let us discuss in detail the morphological differences in the context of plasmoid formation and motion in aforementioned different cases. 

\subsection{Morphological Differences in Multiple Plasmoid Formation and their Motions}

[I] NEL:- In absence of any energy loss effects, plasmoid becomes visible for the first instance around 757 sec which then grows little bigger during its downward propagation. Another plasmoid forms around Y = 95 Mm at 793 sec and then propagates upward. At 830 sec, a less pronounced plasmoid is formed around Y = 25 Mm and is subjected to downward propagation. At around Y = 65 Mm, one more plasmoid forms around 926 sec which further moves downward along the CS. 
\newline
[II] RC only:- In presence of radiative cooling only, the first plasmoid forms around 781 sec at around Y = 65 Mm and undergoes downward motion. Around 818 sec, another plasmoid becomes visible around Y = 100 Mm and starts propagating upward with getting bigger in its dimensions. A very tiny plasmoid is found to move downward starting from around Y = 55 Mm at 926 sec. 
\newline
[III] TC only:- When field-aligned thermal conduction is the only energy loss effect present in our simulation, we found that the first plasmoid forms at Y = 75 Mm around 721 sec and then grows bigger in its voyage towards upward direction. Two less prominent tiny plasmoids are formed around Y = 40 Mm and Y = 90 Mm at 914 sec and 926 sec respectively, These two plasmoids undergo downward and upward propagation respectively. 
\newline
[IV] TC+RC:- In presence of both TC and RC, plasmoid is detected for the first time at 757 sec around Y= 50 Mm and propagates downward with time and grows bigger in dimensions simultaneously. Around 842 s, very less pronounced plasmoid forms close to upward propagating outward plasma and merge with the outflowing plasma instantaneously. No more plasmoids are seen to form within the CS. 

All of these discussed morphological differences are evident in the animation associated with Figure~\ref{label 2}. Therefore, even though, the onset of CS dynamics has been carried out via exact same prescription as proposed by \citet{2024ApJ...963..139M} except asymmetric interaction of velocity perturbation and the CS, the later phase dynamics within the CS is different in all four cases. Animation roughly indicates that the dynamics within the CS may evolve in similar manner for aforementioned four cases upto around 600 s, i.e., before fragmentation of the CS. However, we will first discuss about the thinning and elongation of the CS in all the cases till 600 s as quantitative measures to closely examine the similarities or differences in CS dynamics before focusing on plasmoid formation and its subsequent motions. It is to be noted that we do not consider first 228 s while undertaking any quantitative estimations to ensure that the initial velocity pulse (which perturbs the force-free CS) has completely left the simulation domain without any reflections from all boundaries.

\subsection{Temporal Variation of CS Dimensions before Onset of Fragmentation}
\begin{figure*}
\centerline{\hspace*{0.02\textwidth}
         \includegraphics[width=0.53\textwidth,clip=]{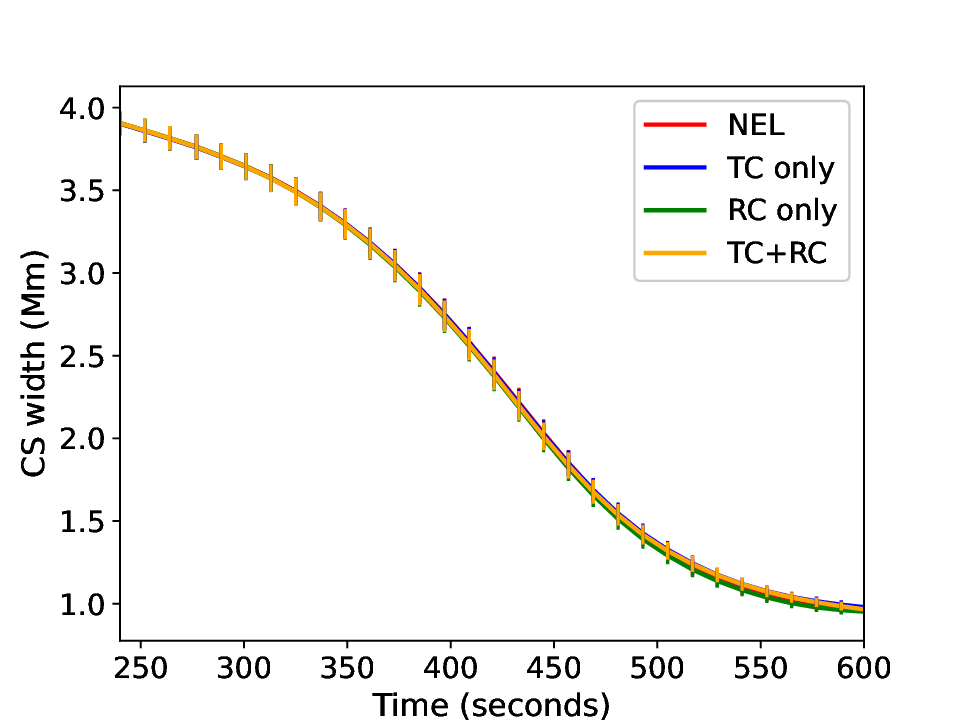}
         \hspace*{-0.03\textwidth}
         \includegraphics[width=0.53\textwidth,clip=]{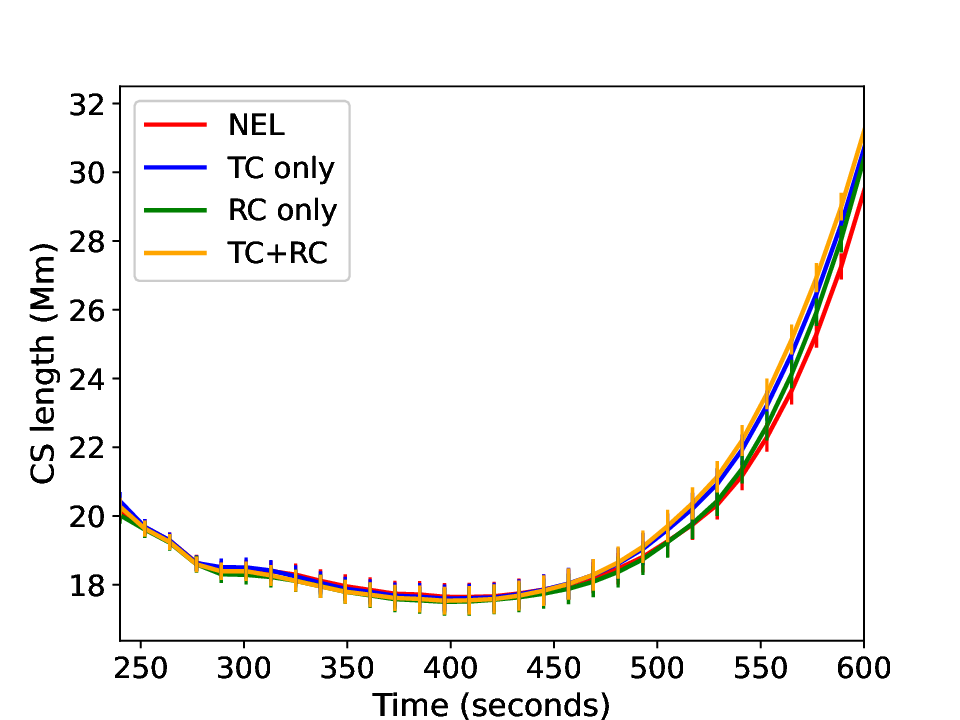}
           }
\vspace{-0.39\textwidth}   
\centerline{ \Large     
\hspace{-0.035 \textwidth}  \color{black}{(a)}
\hspace{0.45\textwidth}  \color{black}{(b)}
   \hfill}
\vspace{0.36\textwidth}    
\caption{Panel (a) and (b) exhibits variation of CS widths and CS lengths with time from 240 s to 600 s. The errorbars are plotted after multiplied by a factor of 2 for better visualization. Panel (a) exhibits that there are no differences between CS widths throughout the considered time window for all the four case studied. The minimal visible distinctions in CS length profiles after around 500 s are not significant enough to be attributed to presence of energy loss effects in the thinning phase.}
\label{label 3}
\end{figure*}

Up to 600 s, CS does not undergo to a visibly detectable fragmentation and further a plasmoid formation stage in any of the considered cases. Therefore, we estimate width and length of the CS as a full width half maximum (FWHM) of Gaussian functions fitted on the current density distribution across the CS at $y=65~\mathrm{Mm}$, and along the CS at $x=0~\mathrm{Mm}$ respectively. Basically we use \citep{2024ApJ...963..139M}
\begin{equation}
   G(s,\sigma) = C~\exp\left(\frac{-(s-\bar{S})^{2}}{2\sigma^{2}}\right),
\end{equation}
where $s$ can either be $x$ or $y$ for estimation of CS width or length respectively. $C$ is the maximum of the current density which is also the peak of the Gaussian function. $\bar{S}$ is the $x$ or $y$ coordinates of the Gaussian peak during estimation of width or length respectively. The estimated value of \(\sigma\) is then used to obtain FWHM of the distribution as
\begin{equation}
   \mathrm{FWHM = 2~\sqrt{2 \sigma^{2}~ln2}}.
\end{equation}
We find that the estimated widths of CS are decreasing at similar rates and similar manner for aforementioned four cases (See panel (a) of Figure~\ref{label 3}). Therefore, we infer that the thinning of the CS entirely depends on the external perturbation and is independent of the presence or absence of cooling effects. On the other hand, initially the lengths of the effective diffusion region along the CS slightly decreased from 240 to 420 s (See panel (b) of Figure~\ref{label 3}). Thereafter, the CS lengths are found to increase at a faster rate after around 420 s which may be a signature of onset of magnetic reconnection (See panel (b) in Figure~\ref{label 3}). There seems to have differences in the measured CS lengths once we reach close to 600 s (See panel (b) in Figure~\ref{label 3}) However, close inspection suggests that  current density profiles along the CS start to deviate slightly from Gaussian profile after around 540 s due to its gradual flattening. This results in an additional systematic error along with random error associated with Gaussian fitting. Errorbars in both panel of Figure~\ref{label 3} stand only for random errors associated with Gaussian fitting and do not consider systematic error due to flattening of the current density profile. Since, flattening and deviation from Gaussian profile will vary with time in all the cases, therefore increasing differences in estimated CS lengths can also be within the total uncertainties of the measurements. Therefore, even though estimated FWHMs in the time window 540-600 s gives an indication of how the CS lengths are changing with time, they are with certain degrees of uncertainnesses for drawing any important physical interpretation about presence of energy loss effects on the basis of differences in CS length in different cases. Therefore, slight deviations are not considered as significant enough to correlate with presence of energy loss effects in presence of case-specific time-varying systematic errors. Also, as discussed in section 3.8 and 3.9, there is no distinction in rate of decrease in average magnetic energy density and increase of average kinetic energy density in the time window of interest, i.e., 480-600 s for different four cases which further suggests that energy redistribution due to presence of energy loss effects are not effective in this time window. Therefore, we conclude that energy loss effects do not play any significant role in CS dynamics till 600 s, i.e., in the thinning and elongation phase of the externally perturbed CS. Now, let us shift our focus on the onset times of fragmentation and deduce Lundquist number and aspect ratio as two indicators of onset of tearing within the CS at those times.
\subsection{Estimation of Instantaneous Lundquist Number and Aspect Ratio at Onset Time of Fragmentation of CS}
As discussed in previous section, we fit the current density profiles along and across the CS with Gaussian functions to find out the length and width of the CS till 600 s. But after 540 s itself, the current density profiles become more flatter along the CS so that they can not be fitted properly using Gaussian function. However, peak of current density along and across the CS should be closely consistent with each other, i.e., the difference is less than and not close to 1 until there is onset of fragmentation of the CS. So, we choose to set a criterion that the time when those two peaks are differing from each other by more than 1 in normalized units will be taken as the onset time of fragmentation. Also, since, we can not rely anymore on FWHM extracted from Gaussian fitting to deduce CS length, we rather find out the distances along CS at which the current density becomes half of the instantaneous maximum. We consider the difference between pair of such distances as the length of the CS. Also, we find out the average Alfv\'en speeds in between those pair of distances. In this way, we find that in case without energy loss, fragmentation starts at 661 s with estimated CS length being 50.79 Mm. Instantaneous average Alfv\'en speed is found to be  $310~\mathrm{km.s^{-1}}$. Therefore Lundquist number is equal to $6.56 \times 10^{4}$ at the onset of fragmentation or tearing along the CS in absence of any energy loss effects.  When only TC is present as energy loss effect, fragmentation starts around 625 s within a CS of 39.53 Mm length. Average Alfv\'en speed is found to be $359~\mathrm{km.s^{-1}}$. Hence, estimated Lundquist number is $5.91 \times 10^{4}$. On contrary, fragmentation starts later at 685 s when only RC is present. In that case, CS length and average Alfv\'en speed are estimated to be 62.19 Mm and $291~\mathrm{km.s^{-1}}$ which corresponds to a value of $7.54 \times 10^{4}$ of Lundquist number. In similar way, we find that when both TC and RC are taken into account, the onset of fragmentation takes place around 661 s within a CS having 51.42 Mm length. Average Alfv\'en speed is found to be $319~\mathrm{km.s^{-1}}$. This corresponds to a Lundquist number of $6.8 \times 10^{4}$. The corresponding aspect ratio of the CS are 51, 41, 61 and 56 for case NEL, TC only, RC only and TC+RC respectively. It has been found that aspect ratio of the CS should be equal or higher than $S_{L}^{1/3}$ for development of ideal tearing instability, where $S_{L}$ stands for Lundquist number \citep{2001EP&S...53..473S,2014ApJ...780L..19P,2016JPlPh..82e5301T}. Here, estimated aspect ratios and instantaneous Lundquist numbers in all four cases validate such criterion. If, we consider aspect ratio is equal to $S_{L}^{\alpha}$ \citep{2009PhPl...16k2102B}, we find $\alpha$ to be 0.35, 0.34, 0.37 and 0.36 for case NEL, TC only, RC only and TC+RC respectively, which are close to but slightly higher than $1/3$. Therefore, the comparison of values of Lundquist number, aspect ratio as well as onset times of fragmentation possibly suggest that presence of thermal conduction is advancing the onset of tearing process, whereas presence of radiative cooling has the opposite effect. We further estimate growth rates of linear tearing instability using \citep{2009PhPl...16k2102B}  
\begin{equation}
\gamma_{L,max} = S_{L}^{(3\alpha-1)/2}\times v_{A}/L
\end{equation}
at the onset time of tearing with L and $v_{A}$ being the instantaneous  CS length and average Alfv\'en speed as mentioned above . The estimated growth rates are $8.1 \times 10^{-3}~\mathrm{s}^{-1}$, $10^{-2}~\mathrm{s}^{-1}$, $8.7 \times 10^{-3}~\mathrm{s}^{-1}$ and $9.6 \times 10^{-3}~\mathrm{s}^{-1}$ for without energy loss effects, TC only, RC only and TC+RC which correspond to 125 s, 100 s, 114 s and 104 s as times required for growth of primary plasmoids in those cases respectively. Therefore, presence of TC certainly enhance the growth rate of tearing mode after its onset. Now the times of onset of fragmentations are roughly 96 s before the first instances of visibility of primary plasmoid in all four cases as far as crude visual inspections are concerned. Therefore, theoretically calculated time required for growth of tearing mode is consistent with the approximate time gap between onset of fragmentation and first visible inspection of plasmoids. So, different Lundquist number and aspect ratio required for plasmoid formation suggest that presence of energy loss effect certainly start to affect the CS dynamics once it advances towards tearing instability. Therefore, let us see how the rate of reconnection is getting affected due to presence of such effects in next section.  

\subsection{Temporal Variation of Reconnection Rate from 240 to 962 s}
\begin{figure}
\centerline{\hspace*{0.02\textwidth}
         \includegraphics[width=0.53\textwidth,clip=]{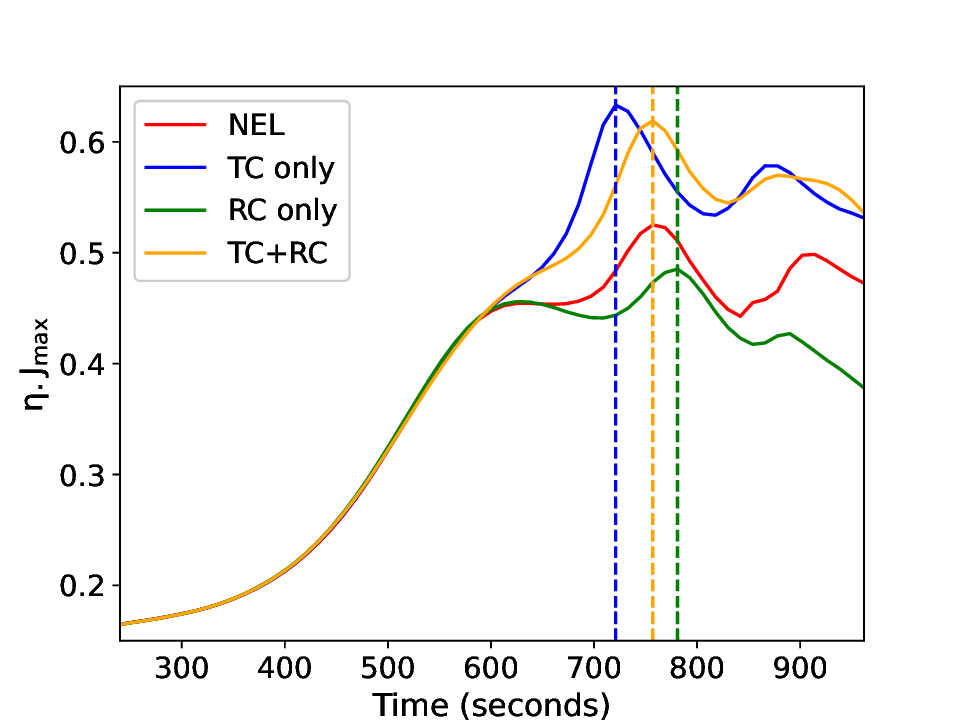}
         }
\caption{Temporal variation of estimated reconnection rate using $\eta J_{max}$ during the period of 240 s to 962 s i.e., upto the end of simulation. Figure gives an indication that the reconnection rate profiles in all the cases are overlapping with each other upto 600 s. After 600 s, reconnection rates keep on increasing before attaining first peak for the cases TC only and TC+RC. On the other hand, for case without energy loss and RC only, reconnection rates become almost constant or decrease slightly before finally subjected to increment just before reaching the first peaks. Blue, orange and green dashed vertical lines denote the times at which primary plasmoids become visible in case with TC only, TC+RC and RC only. Red dashed line is overlapping with the orange one.}
\label{label 4}
\end{figure}

To understand reconnection dynamics, one important way is to measure reconnection rate. We find out the maximum of current density, i.e., $J_{max}$ along the CS and use $\eta J_{max}$ as the measure of reconnection rate following \citet{2001ApJ...549.1160Y,2024ApJ...963..139M}. $\eta J_{max}$ basically gives the resistive component of electric field in Ohm's law. From Figure~\ref{label 4}, it seems that reconnection rate is varying in similar way for all cases up to 600 s without any detectable differences. After 600 s, reconnection rate keeps on increasing in presence of TC only and TC+RC before reaching peaks at 721 s and 757 s respectively. These times are consistent with the instances when the first plasmoids become visible in these two cases (See blue and orange dashed vertical lines in Figure~\ref{label 4}). On contrary, the rate profiles are subjected to the formation of plateaus or nearly constant values in absence of any energy loss mechanism (i.e., NEL case) and RC only case before peaking at 757 s and 781 s respectively during first visualization of their primary plasmoids. After achieving the first peak, reconnection rate exhibits distinct oscillatory behaviour at later stage that is associated to collective outcome of different plasmoid dynamics in those cases. It is evident that presence of TC enhances the reconnection rate, i.e., results in a faster reconnection after onset of tearing instability. On contrary, in tearing unstable CS, reconnection takes place with a slower rate in presence of radiative cooling only than that in case without any energy loss effects. When both the effects are at work, the resultant reconnection rate is faster than that in the case without energy loss and closer to the profile for TC only case. This suggests that energy loss may be taking place mostly through TC even when both TC and RC are present in our system since TC and RC have opposite effects individually. Since, presence of energy loss effects are affecting the rate of reconnection, we then focus to find out dynamical and thermal properties of the plasmoids and reconnection outflows in all four cases as discussed in the following section.  
\subsection{Comparison of Distribution of Temperature and Velocities of Outward Moving Plasmoids along CS}
\begin{figure*}
\centerline{\hspace*{0.02\textwidth}
         \includegraphics[width=0.6\textwidth,clip=]{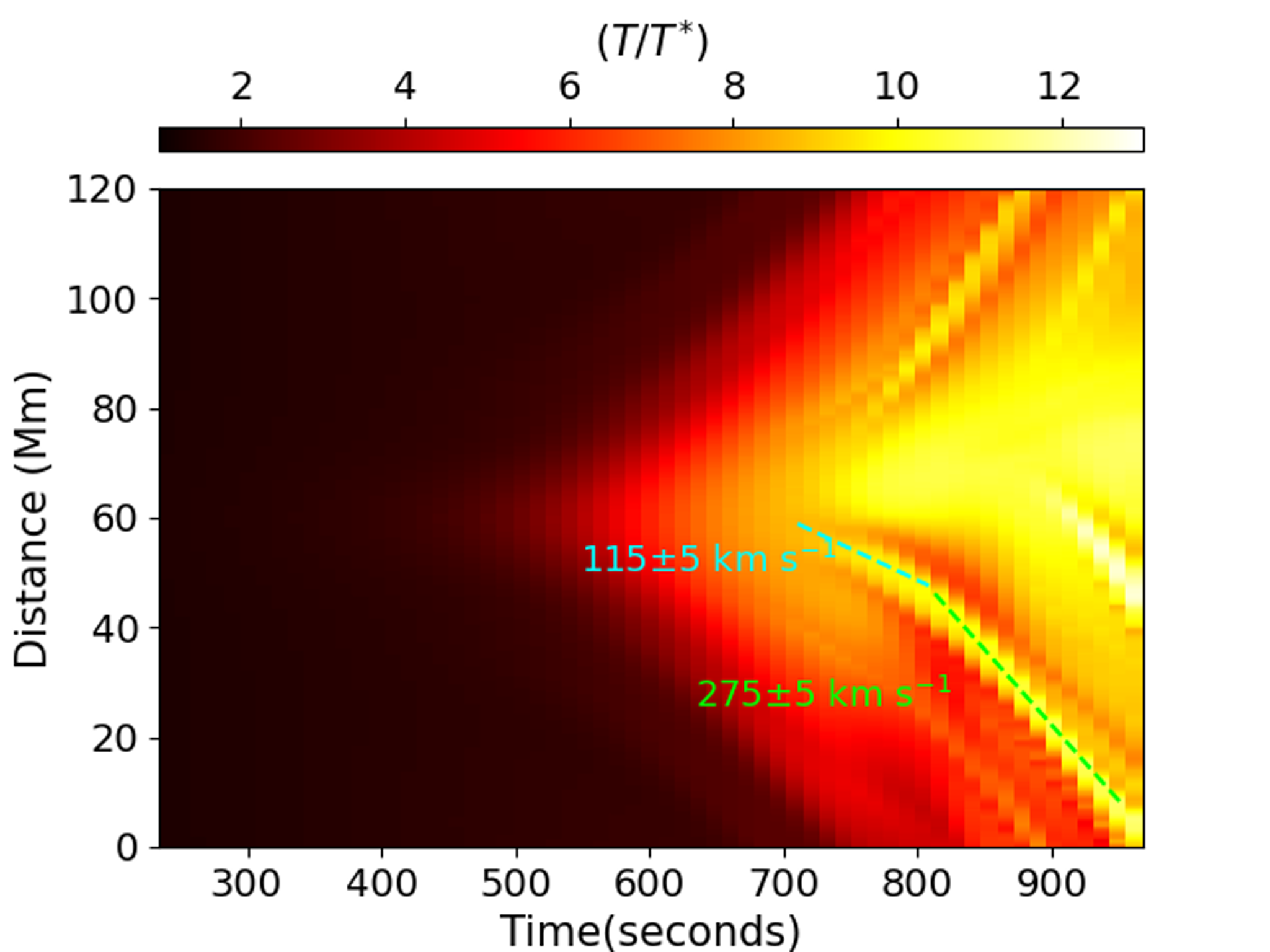}
         \hspace*{-0.03\textwidth}
         \includegraphics[width=0.6\textwidth,clip=]{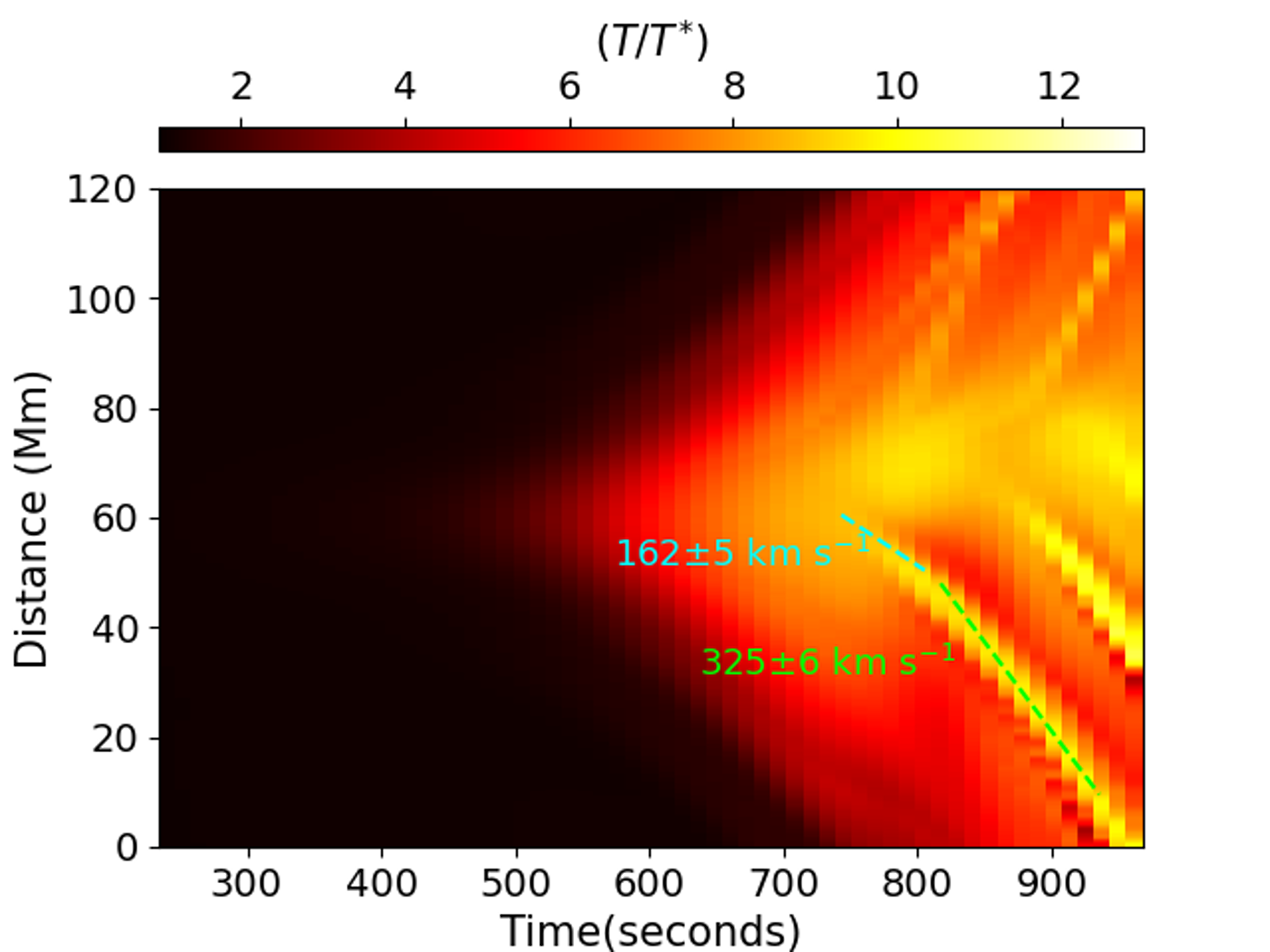}
        }
\vspace{-0.35\textwidth}   
\centerline{ \scriptsize \Large     

\hspace{-0.01\textwidth}  \color{white}{(a) NEL}
\hspace{0.46\textwidth}  \color{white}{(b) RC only}
   \hfill}
\vspace{0.36\textwidth}    
\centerline{\hspace*{0.02\textwidth}
         \includegraphics[width=0.6\textwidth,clip=]{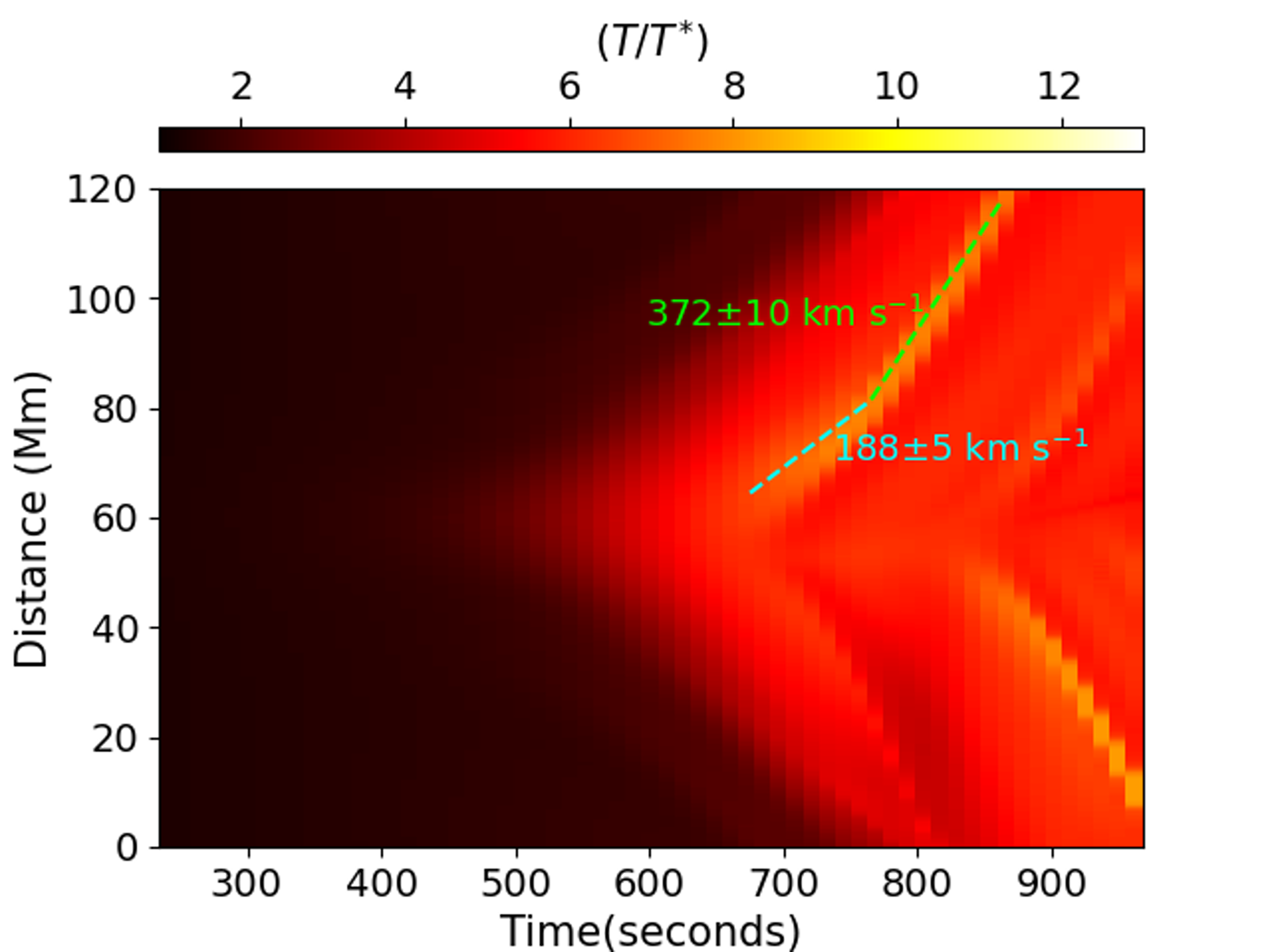}
         \hspace*{-0.03\textwidth}
         \includegraphics[width=0.6\textwidth,clip=]{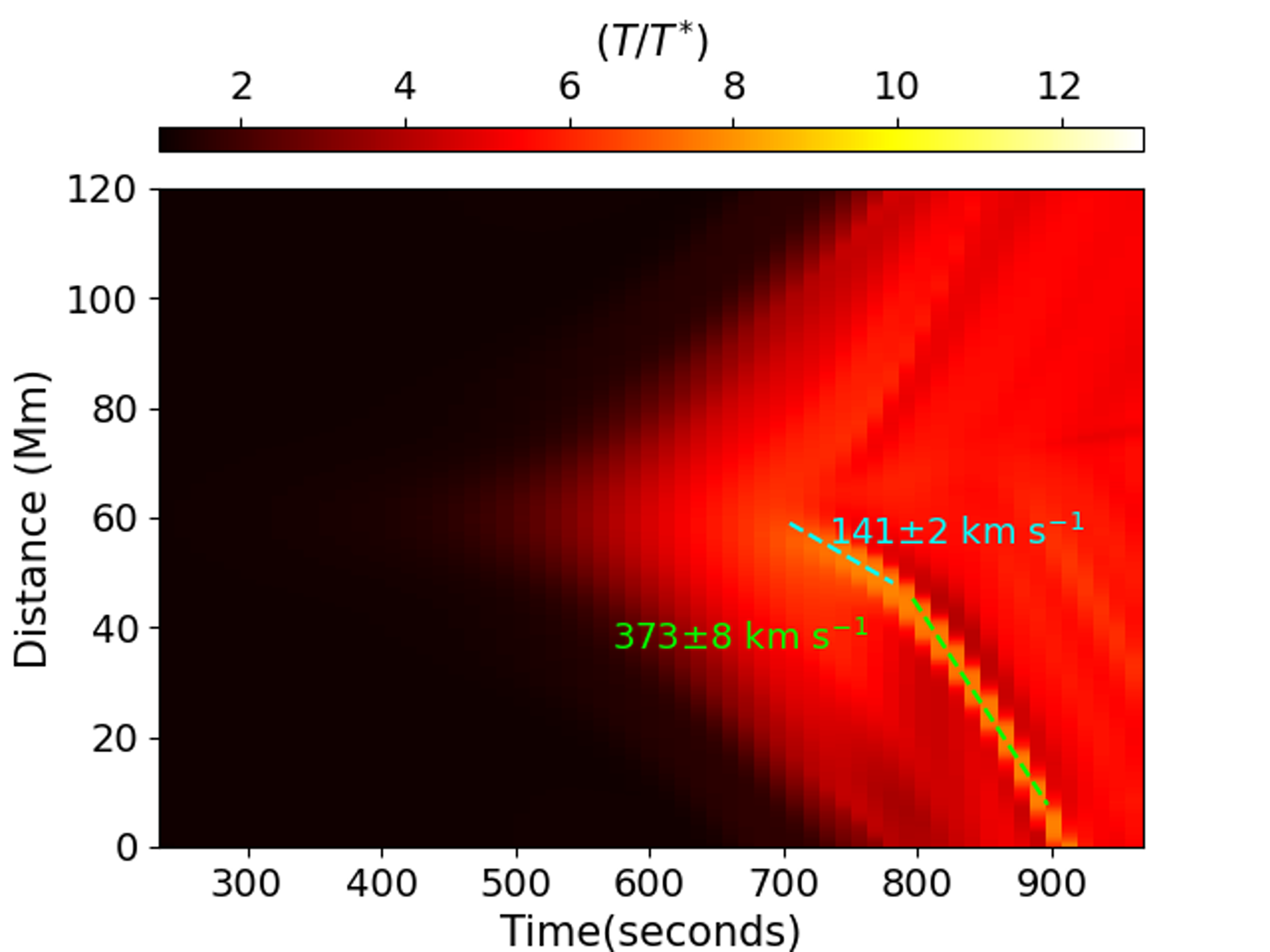}
        }
\vspace{-0.35\textwidth}   
\centerline{\scriptsize \Large     
\hspace{0.01 \textwidth} \color{white}{(c) TC only}
\hspace{0.42\textwidth}  \color{white}{(d) TC+RC}
   \hfill}
\vspace{0.32\textwidth}    
\caption{Reconnection outflows, plasmoid dynamics and its thermal properties are shown for no energy loss (NEL) (panel (a)), only RC (panel (b)), only TC (panel (c)) and both in presence of RC+TC (panel (d)). Spatio-temporal variation of temperature along the CS is derived along $x =0~\mathrm{Mm}$ from $y=5~\mathrm{Mm}$ to $y=125~\mathrm{Mm}$. The start and end point of the slit is taken to be symmetrically placed with respect to 65 Mm, i.e., the location at which the initial Gaussian pulse interacts with the CS first. The accelerated motion of plasmoids are also evident from broken fitted straight lines (cyan and lime-dashed lines). It is to be noted that distances exhibited in the y-axis in all four panels are distances measured from $y=5~\mathrm{Mm}$, i.e., 0 Mm in y-axis corresponds to  $y=5~\mathrm{Mm}$.}
\label{label 5}
\end{figure*}

TC transports heat from hotter to cooler parts of a system via the collisions of constituents of the system itself. On the other hand, RC heavily depends on density of the emitter. So, it will be interesting to examine how the temperature varies along the CS with time in presence of such energy loss mechanisms. Since TC transports heat via transfer of energy due to collisions of the vibrating constituents, presence of TC also result in higher speed of bulk reconnection outflows to transport energy with faster rate. To check whether these physical intuitions are correct or not, we quantify time-distance diagrams in temperature to investigate the spatio-temporal variation in temperature along the CS between $y = 5~\mathrm{Mm}$ and $y = 125~\mathrm{Mm}$ to maintain symmetry about primary interaction region, i.e., $y = 65~\mathrm{Mm}$, which also provide evidence of reconnection outflows and the propagation of multiple plasmoids along the CS. We find that there is hardly any difference in the starting time of reconnection outflows (around 480-500 s) in all four cases under consideration. Figure~\ref{label 5}(a) exhibits higher temperature in regions between distance of 60 Mm to 80 Mm, i.e., $y = 65~\mathrm{Mm}$ to $y = 85~\mathrm{Mm}$ and relatively lower temperature at the outer portions of the CS in absence of any energy loss mechanisms. In presence of RC only, the temperature seems to be reduced slightly in the aforementioned middle portion of CS compared to the NEL case (See Figure~\ref{label 5}(b)). On contrary, in presence of TC or TC+RC, there are no such relatively higher temperature regions visible along the CS (See panel (c) and (d) of Figure~\ref{label 5}). Possibly, TC is dominantly redistributing the energy along the field lines in the outflowing plasma, so it appears more uniform in temperature without any high temperature plasma strands as seen in Figure~\ref{label 5}(a) and (b). The time-distance diagrams also exhibit differences in number of visible plasmoids and their motions, i.e., whether they are moving upward or downward. These notions are consistent with the morphological differences described in Section 3.1.  Since, we find clear evidence of outward plasmoid motions along the CS, we further aim to find out the velocities of plasmoids. We find acceleration profiles for outward moving plasmoids in all the cases. In absence of any energy loss mechanisms, plasmoids are found to have velocity around $115\pm5~\mathrm{km.s^{-1}}$ near to distance of 60 Mm (hereafter, the central region of the CS) which then reaches to $275\pm5~\mathrm{km.s^{-1}}$  during its downward propagation (See Figure~\ref{label 5}(a)). In presence of RC only, the velocity of the plasmoid increases to $325\pm6~\mathrm{km.s^{-1}}$ at outer portion from $162\pm5~\mathrm{km.s^{-1}}$ in the central region (See panel (b) of Figure~\ref{label 5}). In case of TC only and TC+RC, the velocity of the plasmoid increases to $372\pm10~\mathrm{km.s^{-1}}$ and $373\pm8~\mathrm{km.s^{-1}}$ from $188\pm5~\mathrm{km.s^{-1}}$ and $141\pm2~\mathrm{km.s^{-1}}$ respectively (See panel (c) and (d) of Figure~\ref{label 5}). One standard deviations of the measured slopes via straight line fitting are considered as uncertainties of measurements here. Notable increases in the velocities of plasmoids in presence of TC and TC+RC relative to those in absence of any energy loss effect confirm the positive effect of those effects on dynamics within tearing unstable CS at later phases in the magnetic reconnection. Actually in presence of TC and TC+RC, reconnection occurs at a higher rate once the CS undergoes fragmentation (see Figure~\ref{label 4}). Therefore, speed of the reconnection outflows will be higher which further push the plasmoids more to move outward resulting in its higher speed in presence of TC and TC+RC. Now we will discuss the average behaviour of physical quantities like density, temperature along with magnetic and kinetic counterparts of energy densities within the reconnecting CS.
\subsection{Temporal Variation of Average Density from 240 to 962 s}
Density is one important physical quantity of any plasma systems which controls the intensity of radiation emitted from such systems at specific temperatures. Therefore, a knowledge of average density will be a ready reference to understand how much intense the modeled reconnecting CS should be in observational perspective. Also, it will provide a rough idea about the filling factor of high dense plasma within the CS in comparison to the background corona. Since the dynamics along the CS is kinematically and thermally different in four cases, we perform averaging of physical quantities along the CS to find out the effect of presence or absence of energy loss effects on the energetics associated with the dynamic CS undergoing to a tearing instability and plasmoid formation. Since reconnection outflow starts around 480-500 s (see all panels of Figure~\ref{label 5}) and CS width becomes less than 2 Mm after 450 s (see Figure~\ref{label 3}(a)), we have to consider the x extents of the domain used for averaging in such a way that it will cover the entire width of the CS without taking significant contribution from the background in the neighborhood of the CS. Therefore, we perform the averaging of the physical parameters within x = [-1 Mm, 1 Mm] and y= [0, 200 Mm]. From Figure~\ref{label 6}(a), it is evident that average density within the aforementioned domain varies similarly for all the four cases roughly till 588-600 s. After 600 s, there is clear distinction between the estimated average densities for different cases. Average density increases at higher rates in presence of TC only and TC+RC. It reaches 1.75 and 1.84 times the initial value for TC only and TC+RC respectively at the end of the simulation (See Figure~\ref{label 6}(a)). Since, there are less prominent plasmoid formations after primary stage and therefore less secondary fragmentations, the average density keeps on increasing in these two cases. On contrary, in absence of energy loss effects or in presence of RC only, multiple stages of plasmoid formations are present which may create voids of densities in its vicinity which then results in lower values on averaging even in presence of highly dense plasma blobs. So, the average densities are subjected to less increment rate in case of NEL and RC only and reach around 1.27 and 1.33 times of the initial value respectively by the end of the simulation (See Figure~\ref{label 6}(a)).   

\subsection{Temporal Variation of Average Temperature from 240 to 962 s}
\begin{figure*}        
\centerline{\hspace*{0.02\textwidth}
         \includegraphics[width=0.60\textwidth,clip=]{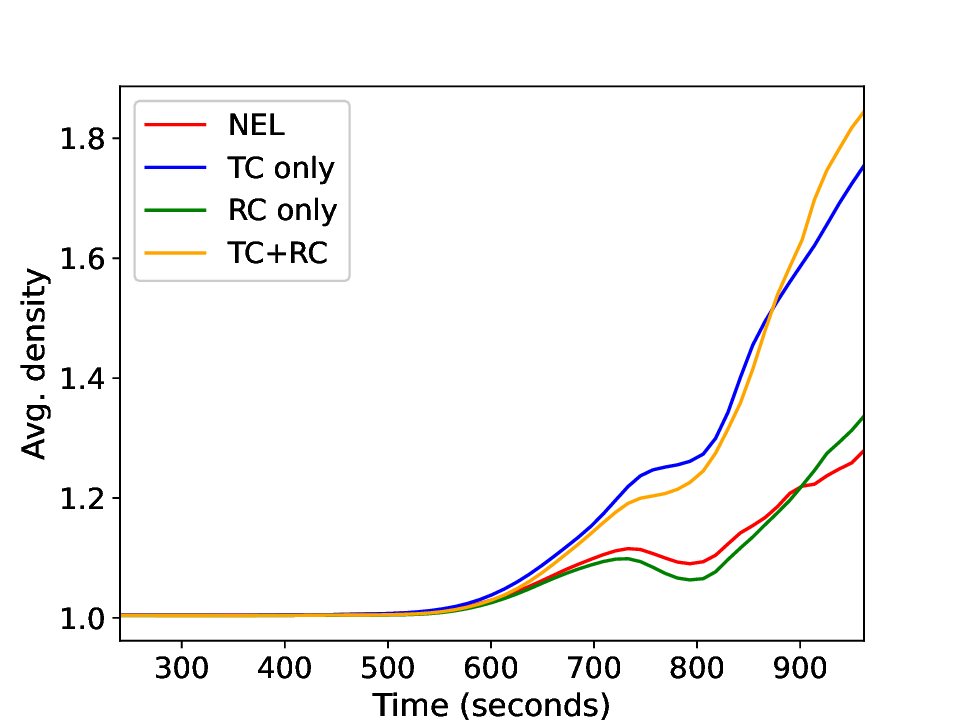}
         \hspace*{-0.03\textwidth}
         \includegraphics[width=0.60\textwidth,clip=]{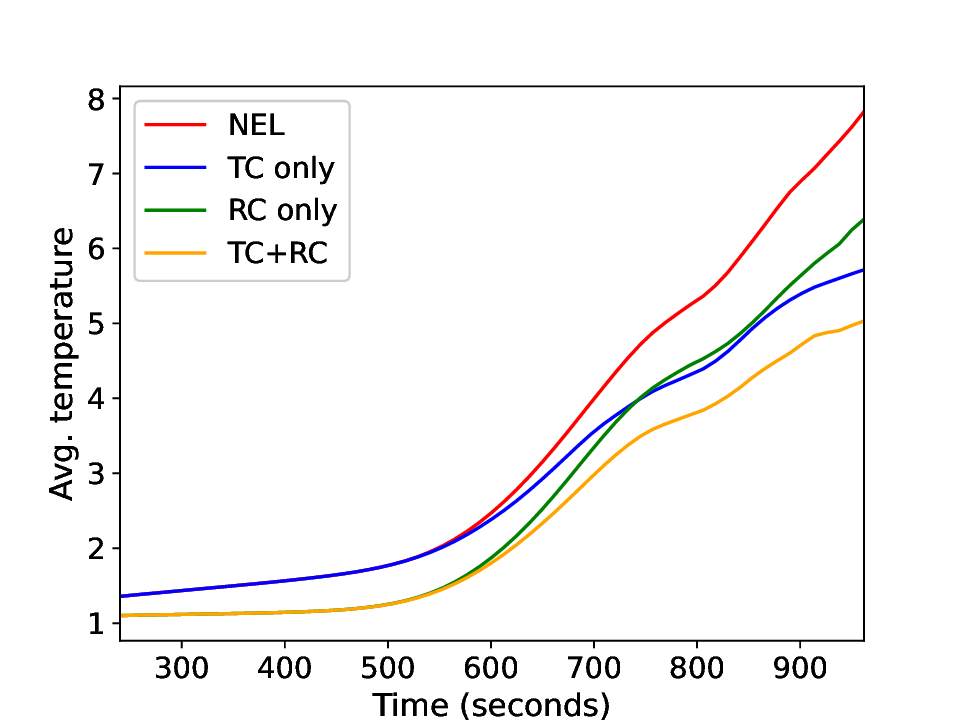}
        }
\vspace{-0.42\textwidth}   
\centerline{\Large    
\hspace{-0.08 \textwidth} \color{black}{(a)}
\hspace{0.52\textwidth}  \color{black}{(b)}
   \hfill}
\vspace{0.40\textwidth}    
\caption{Panel (a) exhibits the variation of average density along the CS with time in dimensionless forms for all four cases. Panel (b) shows the same for temperature. The average has been done within x = [-1 Mm, 1 Mm] and y = [0, 200 Mm]}.
\label{label 6}
\end{figure*}

Even though from Figure~\ref{label 5}, we obtain an approximate idea about the distribution of the temperature along the CS, it will be important to see how the average temperature is evolving with time within the CS to understand the role of energy loss effects on the rate of rise of temperature of the CS as a collective system. Further, it will provide important information about the potential contribution of such reconnection in coronal heating in localized solar corona in presence of energy loss effects. Therefore, we extract the average temperature within the spatial domain denoted in section 3.6. It is found that average temperature keeps on increasing in all the cases (See Figure~\ref{label 6}(b)). However, as expected, presence of energy loss effects certainly decrease the rate of increase of temperature. But the resultant increase indicates that the heating rate must be higher than the corresponding energy loss rates. Nevertheless, the average temperature increases almost 7.8 times of its initial value in absence of any energy loss effects. On contrary, average temperature increases up to 5 times of its initial value when both TC and RC are present as energy loss mechanisms (See Figure~\ref{label 6}(b)). Interestingly, a clear distinction is found in temperature profiles from 240 s itself. Actually, it is evident that presence of RC and TC+RC results in similar average temperature within the CS up to around 588-600 s (See Figure~\ref{label 6}(b)). However, the presence of TC hardly has any effect till 588-600 s as evident from an overlap of temperature profiles without energy loss or with TC only. This scenario indicate that radiative cooling is effective even before the fragmentation of the CS due to enhancement in density in outflow regions. On contrary, thermal conduction comes into play after 600 s. Possible reason behind the increased effectiveness of thermal conduction might be the temperature gradients resulted during fragmentation of the CS which further increases during and after plasmoid formations. However, since there is no prominent secondary fragmentations, the gradients in temperature will not remain high enough for continuous energy loss via thermal conduction. Therefore, the average temperature increases up to around 5.7 times in presence of TC only by the end of the simulation. On contrary, the radiative cooling rate decreases after 600 s as a collective effect of less increment in density and temperature dependent cooling function. Therefore, the temperature increases around 6.4 times of its initial value at the end of the simulation in the presence of radiative cooling only (See Figure~\ref{label 6}(b)).

\subsection{Temporal Variation of Average Magnetic Energy Density from 240 to 962 s }
In magnetic reconnection, stored magnetic energy gets transformed into other forms of energy such as thermal energy, bulk kinetic energy etc. So, it is interesting to see how the presence of energy loss effects affect the rate of decrease of stored magnetic energy within the reconnecting CS and further compare it with rate of reconnection as discussed in section 3.4. Therefore, we calculate the average magnetic energy density ($\frac{B^{2}}{2}$) within the same spatial domain as mentioned in section 3.6. Variation of estimated average magnetic energy with time suggests that the decrements are taking place at very slower rate till 500 s. After 500 s, the rate of decrement becomes faster which is a signature of onset of magnetic reconnection. As expected, the rates are found to be similar in all the four cases till around 700 s (See Figure~\ref{label 7}(a)). Within 600-700 s, the rate is even faster than before which suggest onset of fragmentation of the CS. Hence, magnetic energy starts to be diminished at a higher rate when the CS undergoes fragmentation. After around 720-733 s, the rates of decrement of average magnetic energy density become distinct in all four cases. In presence of TC and TC+RC, magnetic energy decreases at a faster rate (See blue and orange curves in Figure~\ref{label 7}(a)) which is consistent with higher reconnection rate in those cases as discussed in section 3.4. On contrary, the rate of decrement of magnetic energy becomes slightly slower in case without energy loss and RC only. This is also consistent with reconnection rates for these two cases as shown in Figure~\ref{label 4} and discussed in Section 3.4.

\subsection{Temporal Variation of Average Kinetic Energy Density from 240 to 962 s }
\begin{figure*}        
\centerline{\hspace*{0.02\textwidth}
         \includegraphics[width=0.60\textwidth,clip=]{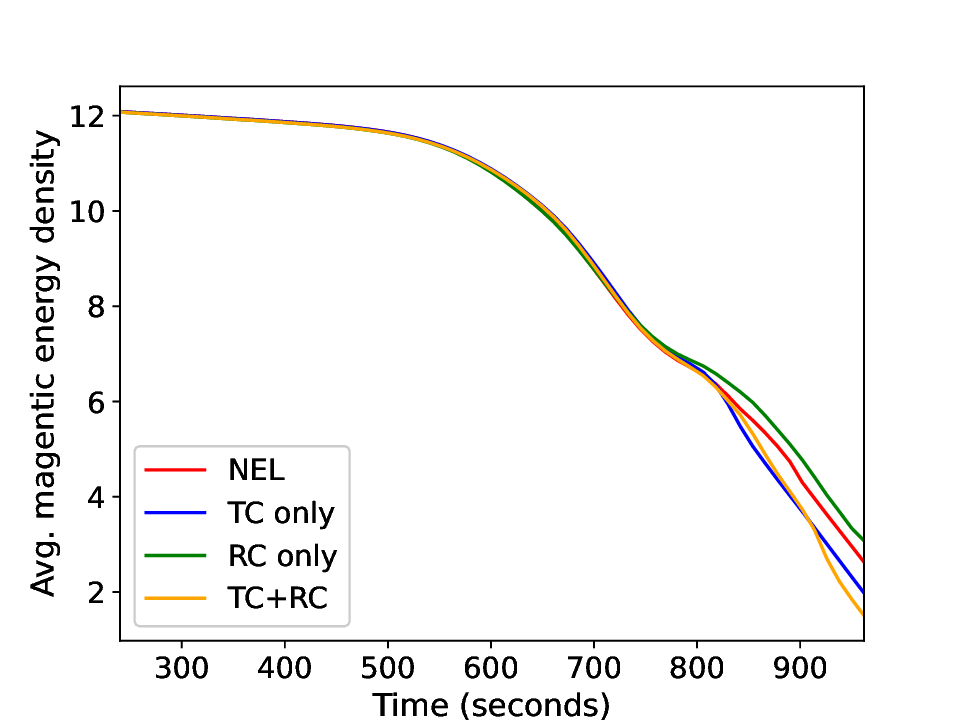}
         \hspace*{-0.03\textwidth}
         \includegraphics[width=0.60\textwidth,clip=]{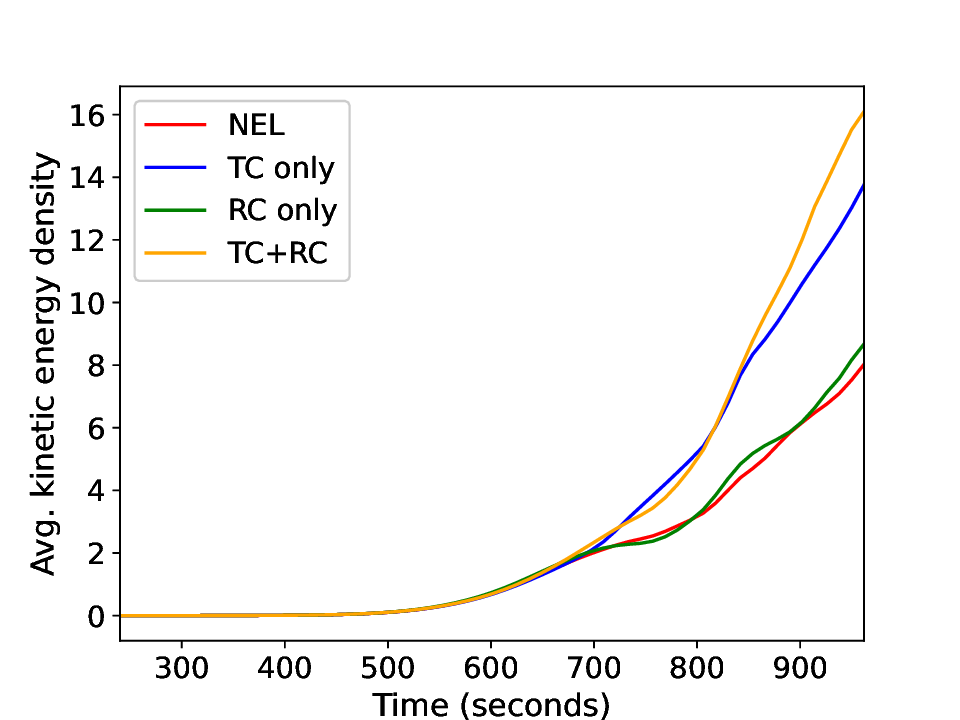}
        }
\vspace{-0.42\textwidth}   
\centerline{\Large    
\hspace{-0.08 \textwidth} \color{black}{(a)}
\hspace{0.52\textwidth}  \color{black}{(b)}
   \hfill}
\vspace{0.40\textwidth}    
\caption{Panel (a) exhibits the average magnetic energy density ($\frac{B^{2}}{2}$) with time for all four cases from 240 s to 962 s. The averaging has been done within x = [-1 Mm, 1 Mm] and y= [0, 200 Mm]. Panel (b) shows variation of average kinetic energy density ($\frac{1}{2}\rho v^{2}$) with time from 240 s to 962 s within same domain along the CS. Both the quantities are shown in dimensionless form.}
\label{label 7}
\end{figure*}

Since magnetic reconnection results in generation of kinetic energy of the plasma embedded within the reconnecting CS, we wish to investigate the differences in rate of increase in kinetic energy with time in all four cases to understand how the energetic outcome of reconnection gets modified due to presence of energy loss effects. Also, since kinetic energy will only become nonzero after onset of reconnection outflows, it will further provide confirmation of the onset time of reconnection. So, estimation of average kinetic energy density ($\frac{1}{2}\rho v^{2}$) is carried out within same spatial domain. Estimated average kinetic energy densities are found to be nearly zero till 500 s (See Figure~\ref{label 7}(b)). This is consistent with the facts that magnetic energies hardly undergoes any decrement till 500 s (See Figure~\ref{label 7}(a)) and reconnection outflows start around 500 s as shown in Figure~\ref{label 5}. From 500 s to 720-733 s roughly, the estimated kinetic energy increases with similar rate in all four cases. This is consistent with similar decrement rates of magnetic energy densities in this time window (as described in previous section). After that, the rate of increment of kinetic energy density becomes higher in presence of TC only and TC+RC (See blue and orange curves in Figure~\ref{label 7}(b)). On contrary, in absence of any energy loss effect and in presence of RC only, kinetic energy density seems to increase with a slightly lower rate with time (See red and green curves in Figure~\ref{label 7}(b)). These higher values of estimated average kinetic energy density in presence of TC only and TC+RC are consistent with the calculated higher speeds of plasmoids in those cases as discussed in Section 3.5.

\section{Discussion \& Conclusion}
In this work, we aim to model an externally driven magnetic reconnection in an elongated coronal CS in absence of any localized enhancement of resistivity due to increase in current density. Following \citet{2024ApJ...963..139M}, we perturb the CS via an external Gaussian velocity pulse. Since, both EUV waves and CSs are ubiquitous in solar corona, it is important to study such kind of interaction. Since, such interaction results in onset of magnetic reconnection in the present case, we consider it as one feature of the Symbiosis of WAves and Reconnection (SWAR) \citep{Sri24, 2024arXiv241102180M}. Basically, here, the system is in magnetohydrostatic equilibrium and does not undergo any dynamics if it is not subjected to wave-like perturbation. So, here, the onset of reconnection is fully attributed to interaction of wave-like perturbation with the CS. Hence, this simulated event qualifies as one example of SWAR.

In the present case, velocity pulse is propagating anisotropically, i.e., the standard deviation in x-direction is higher than that in y-direction. Unlike \citet{2024ApJ...963..139M}, pulse also initially interacts with the CS at an off-centred location along its length. These initial considerations provide a more general modeled scenario of non-symmetric interaction between CS and anisotropically propagating Gaussian pulse akin of EUV waves in solar corona. Also since both thermal conduction and radiative loss are important in solar coronal conditions, it is of utmost importance to include them while studying any magnetoplasma processes such as magnetic reconnection and associated physical conditions in solar corona. Therefore, we also consider such energy loss effects in this present work to examine how they affect various stages of the reconnection dynamics and associated energetics both individually and compositely. Here, we summarize the important scientific findings as follows-
\newline
[i] Before onset of fragmentation, till about 600 s, thinning and elongation of the perturbed CS are nearly independent of presence or absence of energy loss effects.
\newline
[ii] For all the cases, estimated values of local Lundquist number, i.e., $S_{L}$ just at onset of fragmentation are certainly higher than previously reported lower threshold of  $3 \times 10^{4}$ as deduced by \citet{2009PhPl...16k2102B} in their simulation. However, estimated Lundquist numbers are almost 2 times of their reported value. Now, the criterion of aspect ratio being higher than $S_{L}^{1/3}$ is more important for onset of tearing mode instability. Therefore, higher values of Lundquist number are possibly needed to satisfy this criterion for our present simulated dynamics with specific rate of thinning, elongation of the CS, decrement in magnetic energy controlled via our choice of magnetic diffusivity and inflows induced via interaction of velocity perturbation and the CS. 
\newline
[iii] Relative comparison of onset time of fragmentation, Lundquist number and  aspect ratio suggest that TC advances onset of the primary tearing process, whereas presence of RC delays the onset of tearing. We find that the times of first visible detections of plasmoid in all the cases are 96 s after the respective onset times of fragmentation in those cases. 
\newline
[iv] Reconnection is taking place at similar rates in all four cases before the start of fragmentation. But once the CS fragments to form plasmoids subsequently, presence of TC and TC+RC result in faster reconnection than in other two cases.
\newline
[v] Reconnection outflows and plasmoids undergo acceleration in propagation speeds during their outward motion along the CS. Plasmoids finally attain much higher velocity when the energy loss effects are present than in their absence.
\newline
[vi] Presence of TC and TC+RC result in more uniform temperature distribution along the CS than in other two cases. This suggest that thermal conduction is the dominant mechanism of energy loss in solar coronal condition.
\newline
[vii] In tearing unstable CS, the average density increases with higher rate in presence of TC and TC+RC than in other two cases. This confirms more uniform distribution of dense plasma along the CS in those cases.
\newline
[viii] As expected, average temperature increases with respectively highest and lowest rate in absence of any energy loss effects and in presence of both TC and RC simultaneously.
\newline
[ix] When the CS is undergoing tearing and plasmoid formation, average magnetic energy density decreases with higher rates in presence of TC and TC+RC than in other two cases. This is consistent with faster reconnection rate in presence of TC and TC+RC in tearing unstable CS.
\newline
[x] In tearing unstable CS, average kinetic energy density increases with considerably higher rates when TC and TC+RC are present. On contrary, the rates of the same remains lesser in the case without \textbf{energy loss} and in presence of RC only.

\citet{2024ApJ...963..139M} showed that the reconnection dynamics took place as per considered physical magnetic diffusivity instead of being controlled by numerical diffusivity. In their simulation, the smallest grid-size was 97.5 km. Since, in the present work, the smallest grid-size is 78 km, the simulated dynamics is certainly physical instead of being governed by numerical diffusivity. In \citet{2024ApJ...963..139M}, the reconnection rate was found to be less than 0.45. But here, the rate even reaches more than 0.6 in presence of the thermal conduction. Even in NEL case in the present work, the rate is reaching around 0.52 on contrary to 0.45 in \citet{2024ApJ...963..139M}. In \citet{2024ApJ...963..139M}, reconnection rate is subjected to a less oscillatory nature followed by a steep decrease of its value. They have described the presence of monster plasmoid at the centre of the CS as the reason behind that particular physical scenario. In this present case, oscillatory profiles are more prominent, and these rate profiles do not suffer any steep decrement at later times. This feature is consistent with the absence of any steady monster plasmoid.

\citet{2022A&A...666A..28S} reported that tearing can take place even at Lundquist number smaller than $3 \times 10^{4}$ when the system is subjected to thermal imbalance between radiative cooling and time-independent background heating. Here, we do not have any constant background heating term, rather we rely on Ohmic heating only to counterbalance radiative cooling. We notice that in the present case, presence of radiative cooling is delaying the onset of tearing process via requirement of higher Lundquist number and higher aspect ratio. But the growth rate of linear tearing mode after onset of fragmentation increases slightly in presence of radiative cooling in comparison to that in absence of energy loss effects. Basically, \citet{2022A&A...666A..28S} reported that tearing can be initiated at lower Lundquist number in presence of RC than in absence of energy loss mechanisms. On the other hand, we find that higher Lundquist number is required for onset of fragmentation in presence of RC in our study.  However, enhancement of growth rate of tearing mode after its onset in presence of RC in comparison to that in absence of energy loss effects in the present study supports the findings of \citet{2022A&A...666A..28S}. \citet{2022A&A...666A..28S} used magnetic field perturbation to perturb the CS throughout its length simultaneously in absence of any guide field. On contrary, we used wave-like perturbation primarily interacting with the CS at certain location in presence of guide magnetic field. We observe only a few plasmoids instead of having a bead-like feature formed by plasmoid chain along the CS \citep{2022A&A...666A..28S}. So, we do not have large number of highly dense plasma blobs as in \citet{2022A&A...666A..28S} which may result in less-effectiveness of radiative loss in our simulation. 

\citet{2023A&A...678A.132S} studied thermally coupled tearing in a 3D initially force-free coronal CS in presence of background heating, radiative loss and field aligned thermal conduction. They reported that thermal conduction is only trying to homogenize temperature along the CS. On contrary to  \citet{2022A&A...666A..28S}, they found that the CS undergoes a pure tearing instability initially, resulting in 3D topological changes in the magnetic field to form magnetic flux ropes, and only thereafter exhibits runaway condensations near the central portion of the CS. Even though we do not have signature of thermal runaway process in our present simulation, it will be worthy to extend the current study to 3D in future to examine the role of energy loss effects in formation and evolution of flux ropes in 3D setup and make several scientific comparisons since we find that thermal conduction is playing more important role in the present simulated reconnection dynamics than that in \citet{2023A&A...678A.132S} and will transport thermal energy or heat additionally along the guide fields associated with the flux ropes. We will undertake such study in near future. 

Finally, we conclude that the externally driven reconnection dynamics strongly depend on the nature of external perturbation, its interaction point along the CS. Depending on nature of perturbation, the CS undergoes different types of tearing instability, i.e., primary tearing sets at once in either in the entire CS to form chain of small plasmoids or in some specific parts of the CS to generate fewer plasmoids. Also, presence of guide field may suppress tearing of the CS. Mode of tearing instability results in diverse nature of temperature gradient and dense plasma distribution along the CS. Hence, nature of external perturbation and presence of guide field control the effectiveness of thermal conduction and radiative cooling to affect the reconnection dynamics. In this present work, we only consider external forcing aspect to initiate reconnection in presence of uniform resistivity independent of current density and time. Consideration of such enhancement in resistivity might affect the entire reconnection dynamics in terms of how faster it will be. But whether it will alter the effect of presence of energy loss effects, that will be interesting to explore in future work. Now, longer CSs with higher Lundquist number are more prone to tearing such as in solar corona. But in other plasma systems such as solar chromosphere, laboratory scales, Lundquist number will not be as high as in corona. Therefore, it will be interesting to see if similar wave-like perturbation can lead to plasmoid formation there. Moreover, role of thermal conduction and radiative loss on externally driven reconnection dynamics and associated energetics in those magnetoplasma systems will be explored in detail in future works.

\section*{Acknowledgments}
We are grateful to the anonymous reviewer for his/her constructive suggestions which help us to improve the manuscript scientifically. We sincerely thank Prof Eric R. Priest for fruitful primary discussions. We are thankful to open source MPI-AMRVAC 3.0 which provides a user-friendly framework to develop new routines required to simulate our scientific idea. S.M. acknowledges the financial support provided by the Prime Minister's Research Fellowship (PMRF) of India. A.K.S would like to acknowledge the ISRO grant(DS/2B-13012(2)/26/2022-Sec.2) for the support of his scientific research.

\vspace{5mm}
\software{MPI-AMRVAC, Python}




\end{document}